\documentclass[aps,aps,preprint]{revtex4-2}
\usepackage{amsmath,amsthm,amssymb}
\usepackage{graphicx}
\usepackage{caption}
\usepackage{subfigure}

\def \mathbi#1{\textbf{\em #1}} 

\begin{document}

\title{Optical refractive index and spacetime geometry}

\author{Xiong Hui}
\email{xionghui@fudan.edu.cn}
\author{Mengcheng Zhu}
\affiliation{School of Microelectronics, Hubei University\\368 Friendship Avenue, Wuhan 430062, China}

\date{\today}
\vspace{10em}

\begin{abstract}
Classical mechanics and geometrical optics are deeply connected with each other. 
In this work, we generalize the analogy between these two disciplines to relativistic conditions. 
Using this analogy, we are able to make light follow the orbits of massive or massless particles in gravitational field, 
by designing a particular optical medium with a prescribed refractive index profile according to spacetime geometry. 
Furthermore, we build the dictionary that connects optical refractive index and spacetime metric, thus extending $F=ma$ optics to the relativistic limit. 
The results shed light on the equivalence between the dielectric properties of optical media and the geometric structure of spacetime. 
\end{abstract}

\keywords{Fermat's principle; geometrical optics; gradient refractive index (GRIN); gravitational field; metric tensor; spacetime geometry; light precession; gravitational deflection}

\maketitle
\newpage

\tableofcontents
\newpage

\section{Introduction}
Our work has been inspired by a 1986 paper, 
in which the authors pioneered to discover a deep analogy between optics and mechanics based on Fermat's principle \cite{1986_paper}. 
Furthermore, it has been found that many basic laws in optics, such as the scattering theory of inhomogeneous media \cite{optics}, 
or the eikonal equation in curved spacetime that can be thought of as a gradient refractive index (GRIN) medium through Plebanski's constitutive equations \cite{Plebanski}, 
can actually compare with and complement the laws of mechanics. 
In turn, the development of optical theory has given rise to an entirely new cross field \cite{geometric_mechanics}. 
However, from today's perspective, the connection between optics and mechanics established in this article is limited. 
For example, the discussion is limited to Newtonian mechanics and does not contain the relativity situation. 
In addition, photons themselves have no mass, and it is arguable to formally define and relate the energy of light to the mechanical energy of an object. 
Light is relativistic everywhere, but the conservation of mechanical energy is only an approximation in the low-speed limit. 
In the case of high-speed motion, or in the environment of strong gravitational fields, these concepts need to be revised.

In this work, we continue to extend this beautiful analogy to general relativity, in which gravity has been geometrized into spacetime metric, 
by transforming the geometrical structure of spacetime into the refractive index profile of an optical medium. 
Using the perihelion precession problem of Mercury as an example, we design an optical medium in which light propagates along a quasi-elliptical orbit 
with precessing pericenters, in the same way as Mercury evolves around the Sun. The results reproduce the orbit of Mercury quite well, 
which prove to be an effective method of translating the general geometrical structure of spacetime to optical refractive index via the dictionary we have compiled. 
The translation between mechanical and optical languages enables us to study general relativity problems in optical experiments, and vice versa, 
to bring more possibilities to optical design. 

\section{Classical model}
\subsection{Optical medium analogous to Newtonian gravity}\label{sec:Newtonian spacetime}
Due to the symmetry of gravitational fields generated by matter distributed in a finite region, 
it is preferrable to use spherical polar coordinates $(r,\theta,\phi)$ and time coordinate $t$ 
to describe the motion of a test particle drawn by gravity. 
Because the trajectories of the particle are confined to and the same on any plane passing through the gravity center, 
we can limit our discussion to the equatorial plane, where $\theta=\pi/2$ and is a constant. 
Under the assumption of weak gravitational fields and slow velocities, 
the spacetime is nearly flat and the Newtonian motion of a test particle depends only on the curvature of time in spacetime \cite{Schutz}, 
for which the metric can be constructed from the Minkowski metric plus a small correction to the $g_{tt}$ entry: 
\begin{equation}\label{eqn:Newtonian metric}
\mathbi{g}=\left(
\begin{array}{cccc}
-1-2\Delta & 0 & 0 & 0 \\
0 & 1 & 0 & 0 \\
0 & 0 & r^2 & 0 \\
0 & 0 & 0 & r^2 \\
\end{array}
\right)
\end{equation}

where $\Delta=-GM/r$, representing some deviation from flatness in the spacetime that is responsible for generating gravity, 
according to Einstein's equivalence principle. 
In general, the geometrical properties of a spacetime are all encoded in its metric, including the geodesics equation: 
\begin{equation}\label{eqn:geodesics1}
\ddot{x}^\rho+\Gamma_{\mu\nu}^{\rho} \dot{x}^\mu \dot{x}^\nu=0
\end{equation}

in which $\dot{x}:=dx/d\tau$, and $\Gamma_{\mu\nu}^{\rho}$ is the Christoffel symbol determined from the metric: 
\begin{equation}
\Gamma_{\mu\nu}^{\rho}=\frac{1}{2} g^{\rho\lambda}
\left( \frac{\partial g_{\mu\lambda}}{\partial x^\nu}+\frac{\partial g_{\lambda\nu}}{\partial x^\mu} 
-\frac{\partial g_{\mu\nu}}{\partial x^\lambda} \right)
\end{equation}

Given the metric in eq.\ \eqref{eqn:Newtonian metric}, the $\Gamma_{\mu\nu}^{\rho}$ are calculated to have $6$ nonzero components. 
Then we can write the geodesic equations parameterized by proper time $\tau$, 
one for each value of index $\rho$ in $\Gamma_{\mu\nu}^{\rho}$. When $\rho=r$, the equation reads: 
\begin{equation}\label{eqn:geodesics2_r}
\ddot{r}=-\frac{GM}{r^2}\dot{t}^2+r\dot{\phi}^2
\end{equation}

Since the metric in eq.\ \eqref{eqn:Newtonian metric} is independent on $t$ and $\phi$, 
the $p_t$ and $p_\phi$ components of the four momentum $\mathbi{p}$ of the test particle are conserved quantities, 
which lead to the following relations: 
\begin{equation}\label{eqn:p_components}
\begin{split}
p_t&= -\left( 1-\frac{2GM}{r} \right) \dot{t}=-E \\
p_\phi&= r^2\dot{\phi}=J
\end{split}
\end{equation}

The constants $E$ and $J$ have physical significance, which represent the specific energy and angular momentum of the particle measured in coordinates $(t,r,\phi)$. 
$E$ consists of three terms approximately; they are rest mass energy that equals $1$, gravitational potential energy and Galilean kinetic energy: 
\begin{equation}
E=1-\frac{GM}{r}+\frac{1}{\,2\,}\vec{p}^{\ 2}
\end{equation}

where $\vec{p}$ denotes the spatial part of momentum $\mathbi{p}$. 
This result agrees with the facts we know about the motion in central force potential. 
Then we obtain from eq.\ \eqref{eqn:p_components}: 
\begin{equation}
\dot{t}^2=E^2+\frac{4GM}{r}=1+\frac{2GM}{r}+\vec{p}^{\ 2}
\end{equation}

For the nonrelativistic case, usually $|\Delta|\ll 1$ and $\vec{p}^{\ 2}\ll 1$ hold. 
With $\dot{t}^2$ replaced by the above result and $\dot{\phi}$ replaced by $J/r^2$, 
and keeping those terms only to the first order of $\Delta$ and $\vec{p}^{\ 2}$ in eq.\ \eqref{eqn:geodesics2_r}, we have: 
\begin{equation}\label{eqn:geodesics2_r_new}
\ddot{r}=-\frac{GM}{r^2}+\frac{J^2}{r^3}
\end{equation}

Introducing $u=1/r$, and using eq.\ \eqref{eqn:p_components} again, 
we can apply chain rule to re-parameterize the geodesics by $u(\phi)$: 
\begin{equation}\label{eqn:chain rule}
\begin{split}
\frac{d}{d\tau}&= \frac{d}{d\phi}\frac{d\phi}{d\tau}=\frac{J}{r^2}\frac{d}{d\phi} \\
\Rightarrow&\ \frac{dr}{d\tau}=-J\frac{du}{d\phi},\ \frac{d^2r}{d\tau^2}=-J^2u^2 \frac{d^2u}{d\phi^2}
\end{split}
\end{equation}

Then eq.\ \eqref{eqn:geodesics2_r_new} becomes: 
\begin{equation}\label{eqn:geodesics2_u}
\frac{d^2 u}{d\phi^2}+u=\frac{GM}{J^2}
\end{equation}

The solution to eq.\ \eqref{eqn:geodesics2_u} is a family of conic curves: 
\begin{equation}\label{eqn:ellipse}
r=\frac{p}{1-e\cos(\phi-\phi_0)}
\end{equation}

The geometric meanings of $p$ and $e$ are the semilatus rectum and eccentricity of the curve respectively, 
which are constituted with the conserved quantities of motion: 
\begin{equation}\label{eqn:semilatus}
p=\frac{J^2}{GM},\ 
e=\sqrt{1-\frac{2(1-E)p^2}{J^2}}
\end{equation}

If $E$ has not been prescribed, $e$ and $\phi_0$ are integral constants to be determined by the initial conditions $r(\phi_i)$ and $r^\prime(\phi_i)$: 
\begin{equation}\label{eqn:initial_cond0}
\begin{split}
&\left\{
\renewcommand\arraystretch{2}
\begin{array}{ll}
r(\phi_i) & =\dfrac{p}{1-e\cos(\phi_i-\phi_0)} \\
r^\prime(\phi_i) & =-\dfrac{pe \sin(\phi_i-\phi_0)}{\left[ 1-e\cos(\phi_i-\phi_0) \right]^2} \\
\ & =-\dfrac{r^2(\phi_i) e}{p} \sin(\phi_i-\phi_0)
\end{array}\right.
\\[1em]
&\Rightarrow\ 
\left\{
\renewcommand\arraystretch{2}
\begin{array}{ll}
e & =\sqrt{1-\dfrac{2p}{r(\phi_i)}+\dfrac{p^2}{r^2(\phi_i)} \left[ 1+\dfrac{r^{\prime 2}(\phi_i)}{r^2(\phi_i)} \right]} \\
\phi_0 & =\phi_i-\arctan \left( \dfrac{p}{p-r(\phi_i)}\cdot \dfrac{r^\prime(\phi_i)}{r(\phi_i)} \right) \\
\end{array}\right.
\end{split}
\end{equation}

When $0<e<1$, the curve is an ellipse. With this solution, the velocity along the orbit of a test particle is: 
\begin{equation}\label{eqn:velocity_profile1}
v(r)=\sqrt{\frac{2GM}{r}-\frac{GM(1-e^2)}{p}}\propto \sqrt{\frac{1}{r}-\frac{1}{2a}}
\end{equation}

where we define $a=p/(1-e^2)$. One can see immediately that $a$ is nothing else but the length of the semimajor axis of the elliptical orbit. 
From the analogy between geometrical optics and classical mechanics, if we confer the spatial velocity distribution upon the refractive index of a medium, 
then light would follow the same trajectory as the test particle. 
Since $v(r,\phi)$ depends only on $r$, we start from a medium with a spherically symmetric refractive index profile that has the same form as $v(r)$ \cite{cycloid}: 
\begin{equation}\label{eqn:refractive_index1}
n(r)\propto v(r)\ \Rightarrow\ 
n(r)=n_0\sqrt{\frac{1}{r}-\frac{1}{2a}}
\end{equation}

in which $a$ is a constant. 
To examine whether the trajectories of light rays are ellipses as expected, we inspect the length of light path in a general medium: 
\begin{equation}\label{eqn:path_length}
S=\int n(\mathbi{r}) |d\mathbi{r}|=\int n(r,\phi) \sqrt{r^{\prime 2}+r^2} d\phi
\end{equation}

in which $r^\prime:=dr/d\phi$, and $|d\mathbi{r}|$ is the length element of optical path. 
Fermat's principle requires that light path take the one with extremal path length, so the integrand $L=dS/d\phi$, 
after the name optical Lagrangian, needs to obey the Euler-Lagrange (E-L) equation: 
\begin{equation}\label{eqn:EL equation}
\frac{\partial L}{\partial r}-\frac{d}{d\phi}\frac{\partial L}{\partial r^\prime}=0
\end{equation}

\begin{figure}[htp]
\centering
\includegraphics[width=0.5\textwidth]{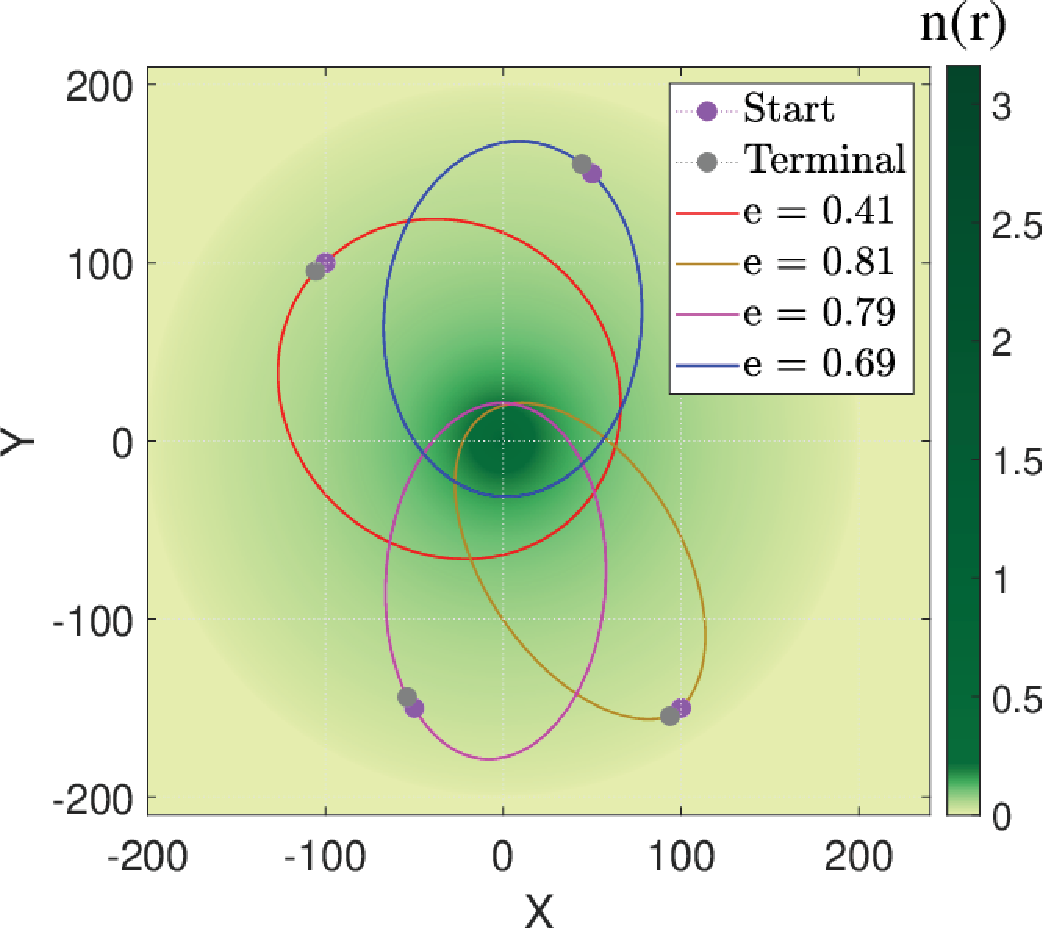}
\caption{Light trajectories analogous to the elliptical orbits of test particles in Newtonian gravity. 
Note that the eccentricity of the orbits changes with the entrance position of light.}
\label{fig:elliptical_orbits}
\end{figure}

From eqs.\ \eqref{eqn:path_length} and \eqref{eqn:EL equation}, we obtain the general form of E-L equation describing stationary light paths 
in spherical polar coordinates: 
\begin{equation}\label{eqn:Binet equation}
r\frac{\partial \ln n(r,\phi)}{\partial r}-\frac{r^\prime}{r}\frac{\partial \ln n(r,\phi)}{\partial \phi}=
\frac{\dfrac{r^{\prime\prime}}{r}-2\left( \dfrac{r^\prime}{r} \right)^2-1}{\left( \dfrac{r^\prime}{r} \right)^2+1}
\end{equation}

Substituting eq.\ \eqref{eqn:refractive_index1} into eq.\ \eqref{eqn:Binet equation}, 
we arrive at the following equation about the orbit $r(\phi)$: 
\begin{equation}\label{eqn:light_orbit1}
\left( 2-\frac{r}{a} \right) r r^{\prime\prime}=\left( 3-\frac{2r}{a} \right) r^{\prime 2}+r^2 \left( 1-\frac{r}{a} \right)
\end{equation}

Setting $w=r^{\prime 2}$, so $2r^{\prime\prime}=dw/dr$, and the above equation can be further simplified to a first order differential equation: 
\begin{equation}
\frac{dw}{dr}=\left( \frac{3}{r}-\frac{1}{r-2a} \right)w+2r+\frac{2ar}{r-2a}
\end{equation}

with the solution
\begin{equation}\label{eqn:r_prime2}
w=\frac{1}{2a}r^2 \left( c_0 r^2-2ac_0 r-2a \right)
\end{equation}

in which $c_0$ is an integral constant. From the expression of $w$, we can further solve $r$ as a function of $\phi$. 
It is found that when $c_0<-2/a$, the solution is: 
\begin{equation}\label{eqn:r_solution}
r=\frac{-\dfrac{2}{c_0}}{1-\sqrt{1+\dfrac{2}{c_0 a}}\cos(\phi-\phi_0)}=\frac{p}{1-e\cos(\phi-\phi_0)}
\end{equation}

This is exactly eq.\ \eqref{eqn:ellipse}, the equation of an ellipse with its semilatus lectum $p$ and eccentricity $e$ defined as: 
\begin{equation}
p=-\frac{2}{c_0},\ e=\sqrt{1+\frac{2}{c_0 a}}=\sqrt{1-\frac{p}{a}}
\end{equation}

The inverse relation gives $a=p/(1-e^2)$, which equals the semimajor axis length of the ellipse as we expect. 
The constant $a$ thus has a geometrical meaning; it implies that once $n(r)$ in eq.\ \eqref{eqn:refractive_index1} has been prescribed, 
all the elliptical orbits should have the same length of semimajor axis. 
In addition, the values of $p$ and $\phi_0$ can be determined from the initial conditions $r(\phi_i)$ and $r^\prime(\phi_i)$. 
Referring to eq.\ \eqref{eqn:initial_cond0}, we solve for $p$ and $\phi_0$, and the result is: 
\begin{equation}\label{eqn:initial_cond1}
\begin{split}
&\left\{
\renewcommand\arraystretch{2}
\begin{array}{ll}
r(\phi_i) & =\dfrac{p}{1-e\cos(\phi_i-\phi_0)} \\
r^\prime(\phi_i) & =-\dfrac{pe \sin(\phi_i-\phi_0)}{\left[ 1-e\cos(\phi_i-\phi_0) \right]^2} \\
\ & =-\dfrac{r^2(\phi_i) e}{p} \sin(\phi_i-\phi_0)
\end{array}\right.
\\[1em]
&\Rightarrow\ 
\left\{
\renewcommand\arraystretch{2}
\begin{array}{ll}
p & =\dfrac{\left[ 2-\dfrac{r(\phi_i)}{a} \right] r^3(\phi_i)}{r^2(\phi_i)+r^{\prime 2}(\phi_i)} \\
\phi_0 & =\phi_i-\arctan \dfrac{r^\prime(\phi_i) p}{r(\phi_i) \left[ p-r(\phi_i) \right]} \\
\ & =\phi_i-\arctan \dfrac{\left[ 2-\dfrac{r(\phi_i)}{a} \right] r(\phi_i) r^\prime(\phi_i)}
{\left[ 1-\dfrac{r(\phi_i)}{a} \right] r^2(\phi_i)-r^{\prime 2}(\phi_i)}
\end{array}\right.
\end{split}
\end{equation}

\begin{figure}[htp]
\centering
\subfigure[Elliptical orbits for a point light source.]{
\includegraphics[width=0.414\textwidth]{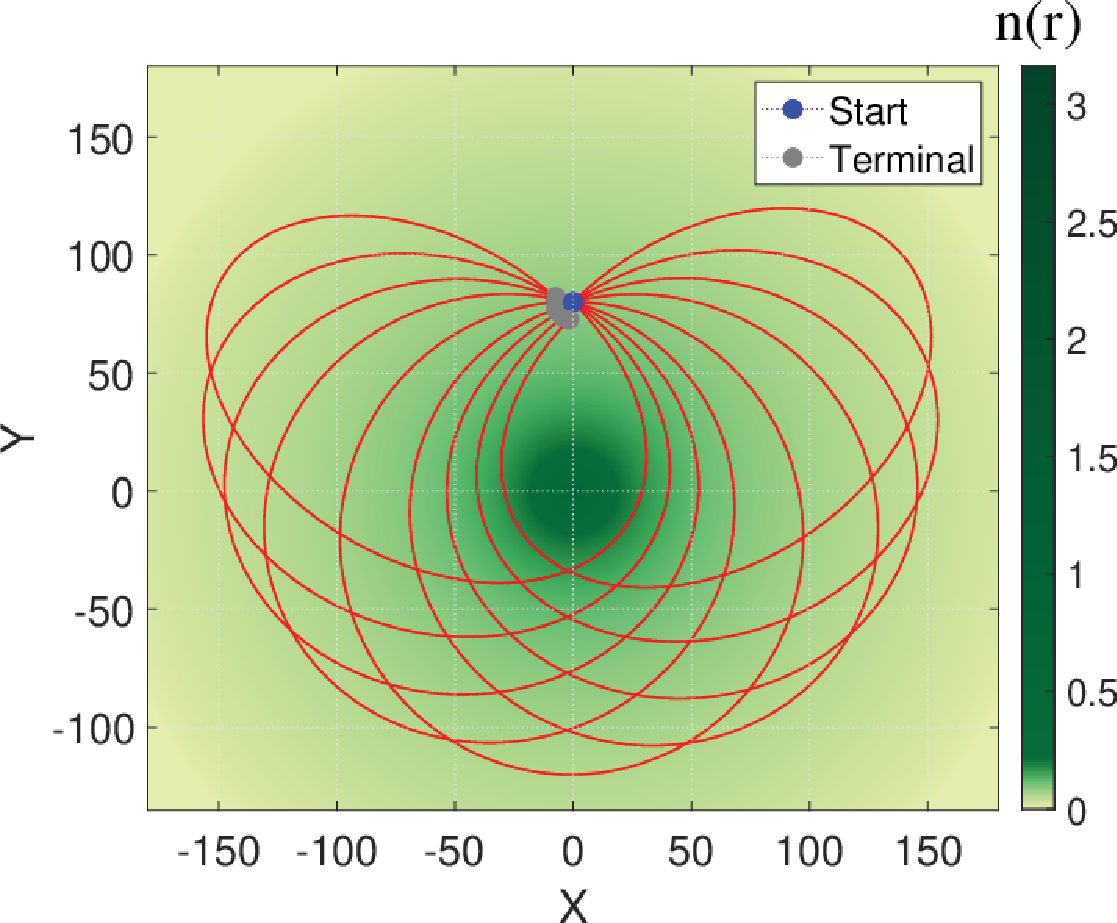}
\label{fig:multi_ellip_orbits_a}}
\quad
\subfigure[Elliptical orbits of parallel light rays.]{
\includegraphics[width=0.4\textwidth]{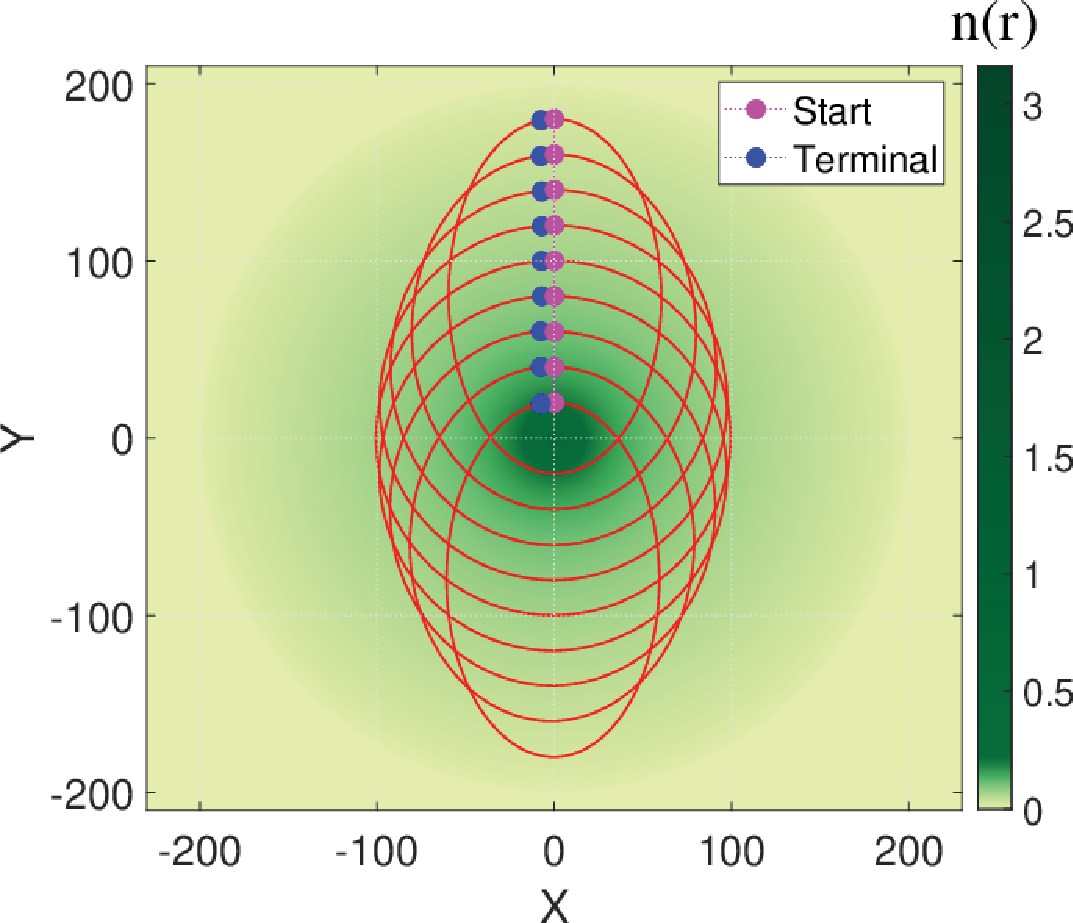}
\label{fig:multi_ellip_orbits_b}}
\\
\caption{(a) Light emits parallely into the gravitational medium. The incident points are arranged along the radial direction. 
As the initial point approaches the medium center, the eccentricity of light orbits decreases at first, down to zero at $r=a$, 
and then increases and becomes elliptical again. The orbital major axes orient radially in the same direction, 
as required by the fact that the center of $n(r)$ lies at the foci of the orbits. 
(b) Light emitting from a point source along different directions and going through elliptical orbits. 
All these trajectories have the same period although their eccentricity is different, so the light rays will return to the initial point at the same time.}
\label{fig:multi_ellip_orbits}
\end{figure}

This completes the solution of light orbits to the E-L equation. 
In numerical simulation, we set $a=100$ and choose different combinations of initial conditions, 
and the calculated light orbits are presented in Fig.\ \ref{fig:elliptical_orbits}. 
On the background mesh colors are used to indicate the value of refractive index. 
It is seen that, light traces strictly follow ellipses with different orientations and eccentricities, depending on the incident position and direction. 
However, all the ellipses have one of their focal points located right at the medium center, 
and the motion of light is just like a planet orbiting around its host star. 
It is worth mentioning that the index profile in eq.\ \eqref{eqn:refractive_index1} constitutes an optical device called the Eaton lens \cite{Eaton}. 

\subsection{Period of elliptical light orbits}\label{sec:ellipse_period}
Next let us look into the period of the light orbit. 
The speed of light measured in the medium is $c/n(r)$, so the time element is $n(\mathbi{r}) |d\mathbi{r}|/c$, 
or the effective optical path length element is $n(\mathbi{r}) |d\mathbi{r}|$. Utilizing eq.\ \eqref{eqn:path_length}, 
we have the integration in spherical polar coordinates: 
\begin{equation}\label{eqn:period_integ}
T=\frac{1}{c}\oint n(\mathbi{r}) |d\mathbi{r}|
=\frac{n_0}{c}\int_{0}^{2\pi}\sqrt{\frac{1}{r}-\frac{1}{2a}}\sqrt{r^{\prime 2}+r^2} d\phi
\end{equation}

With $r^{\prime}$ given by the square root of $w$ in eq.\ \eqref{eqn:r_prime2}, 
the integrand can be recast into the following form: 
\begin{equation}
\begin{split}
T&= \frac{\sqrt{-c_0}}{c}\int_{0}^{2\pi} r\left( 1-\frac{r}{2a} \right) d\phi \\
&= \frac{\sqrt{-c_0}}{c}\left( \int_{0}^{2\pi} r d\phi-\frac{1}{2a} \int_{0}^{2\pi} r^2 d\phi \right)
\end{split}
\end{equation}

There are two parts in the above integration. The first one is: 
\begin{equation}\label{eqn:integral_r}
\int_{0}^{2\pi} r d\phi=-\frac{2}{c_0} \int_{0}^{2\pi} \frac{d\phi}{1-e\cos(\phi-\phi_0)}=-\frac{4\pi}{c_0\sqrt{1-e^2}}
\end{equation}

and the second one is: 
\begin{equation}\label{eqn:integral_r2}
\frac{1}{2a} \int_{0}^{2\pi} r^2 d\phi=-\frac{1-e^2}{c_0} \int_{0}^{2\pi} \frac{d\phi}{\left[ 1-e\cos(\phi-\phi_0) \right]^2}
=-\frac{2\pi}{c_0\sqrt{1-e^2}}
\end{equation}

Then we obtain: 
\begin{equation}\label{eqn:period_Newton}
T=\frac{\sqrt{-c_0}}{c}\left( -\frac{4\pi}{c_0\sqrt{1-e^2}}+\frac{2\pi}{c_0\sqrt{1-e^2}} \right)
=\frac{2\pi}{c\sqrt{(e^2-1)c_0}}=\frac{2\pi}{c}\sqrt{\frac{a}{2}}
\end{equation}

This is a beautiful result, which means that $T$ depends only on $a$, as dictated by $T\propto a^{1/2}$. 
It is the optical version of Keplerian motion of planets in gravitational fields, for which Kepler's third law says $T\propto a^{3/2}$. 
From eq.\ \eqref{eqn:refractive_index1}, we know that a certain refractive index profile $n(r)$ has a certain $a$, 
so that different light orbits inside the medium should all have the same period, 
despite that their locations and shapes may be different. 

This feature makes the medium a competent candidate for designing optical cavities and resonators. 
Due to a constant $T$, light emitting in different directions will return back to the same point after the same amount of time. 
Fig.\ \ref{fig:multi_ellip_orbits_a} shows the light orbits starting from a single point along different directions in the medium with the same $n(r)$ 
as in Fig.\ \ref{fig:elliptical_orbits}. It is evident that the ellipses drawn by these rays are different from each other, 
while the rate at which they propagate forward is highly synchronized. 

\section{Relativistic model}
\subsection{Optical medium analogous to relativistic gravity}
First, let's derive the equation that describes the relativistic motion of a test particle in gravitational fields. 
We start from the Schwarzschild metric that reflects the spherically symmetric geometry of curved spacetime around a static star like the Sun. 
Due to the symmetry of the problem, we select the equator plane to examine the euqation of motion, 
where $\theta=\pi/2$, so the metric tensor $\mathbi{g}$ looks like: 
\begin{equation}\label{eqn:Schwarzschild}
\mathbi{g}=\left(
\begin{array}{cccc}
-\left( 1+2\Delta \right) & 0 & 0 & 0 \\
0 & \left( 1+2\Delta \right)^{-1} & 0 & 0 \\
0 & 0 & r^2 & 0 \\
0 & 0 & 0 & r^2 \\
\end{array}
\right)
\end{equation}

The Christoffel symbols $\Gamma_{\mu\nu}^{\rho}$ under this metric are calculated to have $7$ nonzero components, 
from which we can write the explicit form of the geodesic equation. For the $r$-component, the result is: 
\begin{equation}\label{eqn:geodesics3_r}
\ddot{r}=-\frac{GM}{r^2}\left[ \left( 1-\frac{2GM}{r} \right) \dot{t}^2-\left( 1-\frac{2GM}{r} \right)^{-1} \dot{r}^2 \right]
+r\left( 1-\frac{2GM}{r} \right) \dot{\phi}^2
\end{equation}

Recall from the identity of four velocity $\mathbi{v}\cdot\mathbi{v}=-1$ that: 
\begin{equation}
-\left( 1-\frac{2GM}{r} \right)\dot{t}^2+\left( 1-\frac{2GM}{r} \right)^{-1} \dot{r}^2=-1-r^2 \dot{\phi}^2
\end{equation}

The terms on the left hand side are exactly the expression in the square parenthesis in eq.\ \eqref{eqn:geodesics3_r}, 
so we have by substitution another version of geodesic equation: 
\begin{equation}\label{eqn:geodesics3_r_new}
\ddot{r}=-\frac{GM}{r^2}+r\left( 1-\frac{3GM}{r} \right)\dot{\phi}^2
\end{equation}

In essence, Schwarzschild metric represents the unique solution to Einstein's equation in the static and spherically symmetric case, 
so its components in eq.\ \eqref{eqn:Schwarzschild} are found to be independent on $t$ and $\phi$. 
Following the same reasons as for motion in Newton's gravity, we conclude that there are two conserved quantities, 
the specific energy $p_t=E$ and angular momentum $p_\phi=J$, for the motion in Schwarzschild's spacetime. 
Therefore we refer to eq.\ \eqref{eqn:p_components} again to replace $\dot{\phi}$ by $J/r^2$, 
and then change the derivative $\ddot{r}$ with respect to $\tau$ to that with respect to $\phi$ in eq.\ \eqref{eqn:geodesics3_r_new}. 
The result is: 
\begin{equation}\label{eqn:geodesics3_phi}
\frac{J}{r^2}\frac{d}{d\phi}\left( \frac{J}{r^2}\frac{dr}{d\phi} \right)
=-\frac{GM}{r^2}+\left( 1-\frac{3GM}{r} \right)\frac{J^2}{r^3}
\end{equation}

We further introduce $u=r^{-1}$ and then arrive at the following equation: 
\begin{equation}\label{eqn:geodesics3_u}
\frac{d^2u}{d\phi^2}+u=\frac{GM}{J^2}+3GMu^2.
\end{equation}

Compared to eq.\ \eqref{eqn:geodesics2_u}, there is an additional $u^2$ term that times a small coefficient on the right side. 
Due to the existence of this term, we should find from solving eq.\ \eqref{eqn:geodesics3_u} different geodesics other than standard ellipses. 
Letting $\chi=J^2(GM)^{-1}u$ and $\alpha=3(GM/J)^2$, we obtain a concise dimensionless form of the geodesic equation: 
\begin{equation}\label{eqn:orbit_chi}
\frac{d^2\chi}{d\phi^2}+\chi=1+\alpha \chi^2
\end{equation}

Treating the $\alpha\chi^2$ term as a small correction since $\alpha\ll 1$, this equation can be solved in perburbation method. 
The unperturbed part of eq.\ \eqref{eqn:orbit_chi} is: 
\begin{equation}
\frac{d^2\chi_0}{d\phi^2}+\chi_0=1,
\end{equation}

which is just the Newtonian equation of motion in the unrelativistic case. 
The solution $\chi_0$ is a family of conic curves (to the lowest order of $\alpha$): 
\begin{equation}
\chi_0=1-e\cos(\phi-\phi_0),
\end{equation}

in which $e$ and $\phi_0$ are integration constants. When $0<e<1$, it represents an ellipse, with $e$ being the ellipse's eccentricity. 
This reproduces the result in Newton's gravity, which has been discussed in section \ref{sec:Newtonian spacetime}. 
We then expand the solution $\chi$ according to the order of $\alpha$: 
\begin{equation}
\chi=\chi_0+\alpha \chi_1+\alpha^2 \chi_2+...
\end{equation}

The general solution of $\chi_1$ takes the form: 
\begin{equation}
\chi_1=1+\frac{e^2}{2}-e(\phi-\phi_0) \sin(\phi-\phi_0)-\frac{e^2}{6} \cos2(\phi-\phi_0)
\end{equation}

Therefore we have to the accuracy of first order of $\alpha$: 
\begin{equation}
\begin{split}
\chi&= \chi_0+\alpha \chi_1 \\
&= 1+\alpha-e\left[ \cos(\phi-\phi_0)+\alpha(\phi-\phi_0) \sin(\phi-\phi_0) \right] \\
&\quad\quad +\frac{\alpha e^2}{2} \left[ 1-\frac{1}{3}\cos2(\phi-\phi_0) \right]
\end{split}
\end{equation}

Usually $e\ll1$ holds, so we discard the $e^2$ term on the right hand side, and then $\chi$ reduces to: 
\begin{equation}
\begin{split}
\chi&= 1+\alpha-e\left[ \cos(\phi-\phi_0)+\alpha(\phi-\phi_0) \sin(\phi-\phi_0) \right] \\
&\approx 1-e\cos\left[ (1-\alpha)(\phi-\phi_0) \right]
\end{split}
\end{equation}

This leads us to the well-known solution that describes the orbit with a relativistic feature---
the pericenter precession of non-circular orbits in gravitational fields: 
\begin{equation}\label{eqn:orbit_precess}
r=\frac{p}{1-e\cos\left[ (1-\alpha)(\phi-\phi_0) \right]},
\end{equation}

where $p$ and $e$ have been defined in eq.\ \eqref{eqn:semilatus}. 
With this orbital equation, we further parameterize the orbit by proper time $\tau$ of the test particle. 
The spatial velocity of the particle, relative to a rest frame of the host star system, can be calculated via: 
\begin{equation}\label{eqn:velocity_polar}
\begin{split}
v^2(r,\phi)&= g_{rr} \left( \frac{dr}{d\tau} \right)^2+g_{\phi\phi} \left( \frac{d\phi}{d\tau} \right)^2 \\
&= 
\left[ \left( 1+\frac{2GM}{r} \right) \left( \frac{dr}{d\phi} \right)^2+r^2 \right] \left( \frac{d\phi}{d\tau} \right)^2
\end{split}
\end{equation}

From eqs.\ \eqref{eqn:orbit_precess} and \eqref{eqn:velocity_polar}, we keep terms down to the power of $r^{-2}$ considering $|\Delta|\ll 1$ 
and obtain the expression for $v(r,\phi)$: 
\begin{equation}\label{eqn:velocity_profile2}
\begin{split}
v(r,\phi)&= (1-\alpha) \sqrt{2E-1}\cdot
\left\{ \frac{2GM}{r}-\frac{2(1-E)}{2E-1}+\left[ 1+\frac{3}{4(2E-1)}\cdot \frac{2-\alpha}{(1-\alpha)^2} \right] \frac{(2GM)^2}{r^2} \right\}^{\frac{1}{2}} \\
&\propto \left\{ \frac{1}{r}-\frac{1}{2a(2E-1)}+\left[ 2GM+\frac{p}{2(2E-1)}\cdot \frac{\alpha(2-\alpha)}{(1-\alpha)^2} \right] \frac{1}{r^2} \right\}^{\frac{1}{2}}
\end{split}
\end{equation}

$v(r,\phi)$ turns out to be dependent only on $r$, as it should be due to the spherical symmetry of the problem. 
The correction to the Newtonian case, referring to eq.\ \eqref{eqn:velocity_profile1}, is included in the $p/r^2$ term. 
Compared to the unrelativistic case, the additional $1-\alpha$ factor in eq.\ \eqref{eqn:orbit_precess} is the cause to the pericenter shift of the orbit. 
Between two neighboring pericentric points $(r,\phi_1)$ and $(r,\phi_2)$, the angle difference is: 
\begin{equation}\label{eqn:cycle_angle}
\left| \phi_2-\phi_1 \right|=\frac{2\pi}{1-\alpha}\approx 2\pi(1+\alpha)+O(\alpha^2),
\end{equation}

so the angle of precession per cycle is $\Delta\phi=|\phi_2-\phi_1|-2\pi=2\alpha\pi/(1-\alpha)$. 
A proper value of $\Delta\phi$ can ensure that the orbit is closed, such that the particle will return back to its initial position 
after running multiple cycles of revolution around the gravity center. 

\begin{figure}[htp]
\centering
\subfigure[Orbit in precession with $e=0.5$.]{
\includegraphics[width=0.35\textwidth]{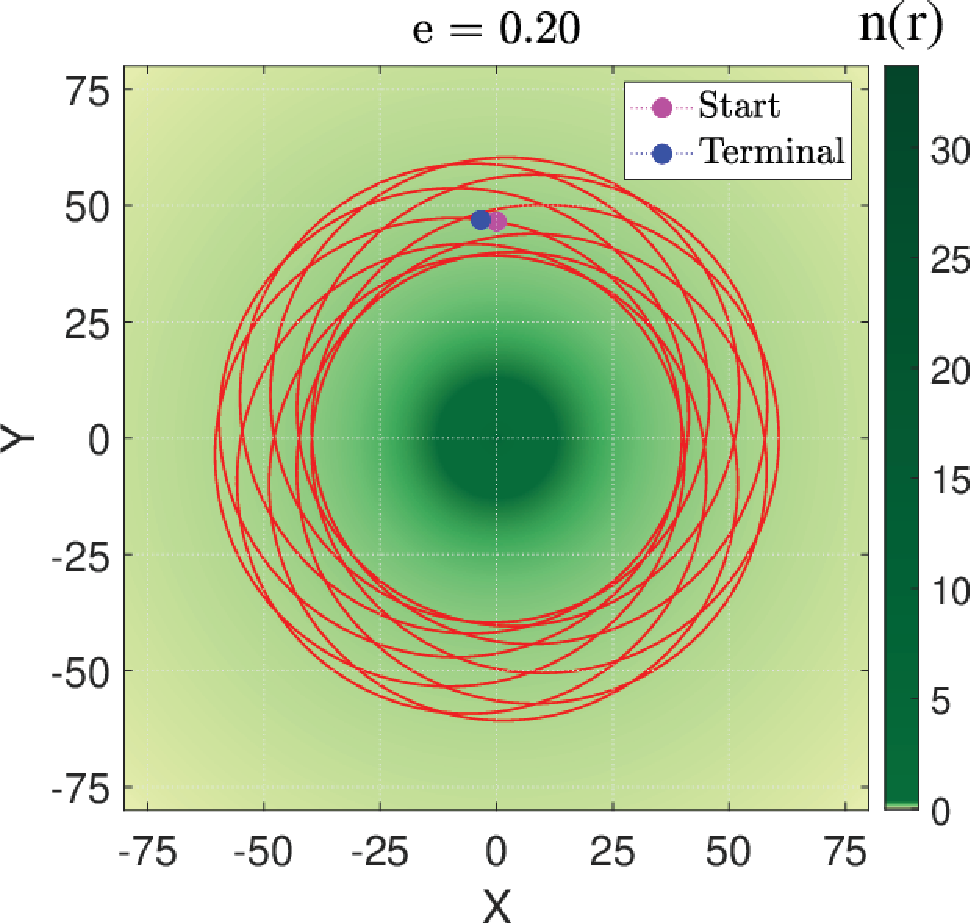}
\label{fig:precess_orbits_a}}
\quad
\subfigure[Orbit in precession with $e=0.6$.]{
\includegraphics[width=0.35\textwidth]{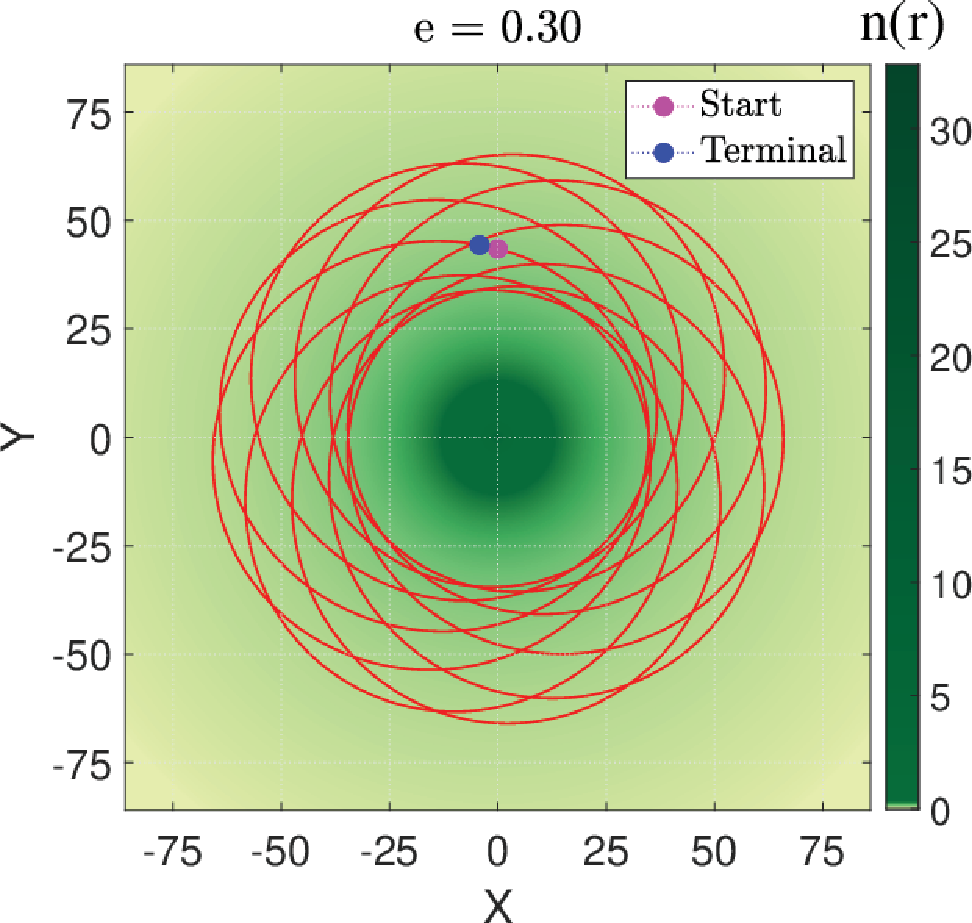}
\label{fig:precess_orbits_b}}
\\
\subfigure[Orbit in precession with $e=0.7$.]{
\includegraphics[width=0.35\textwidth]{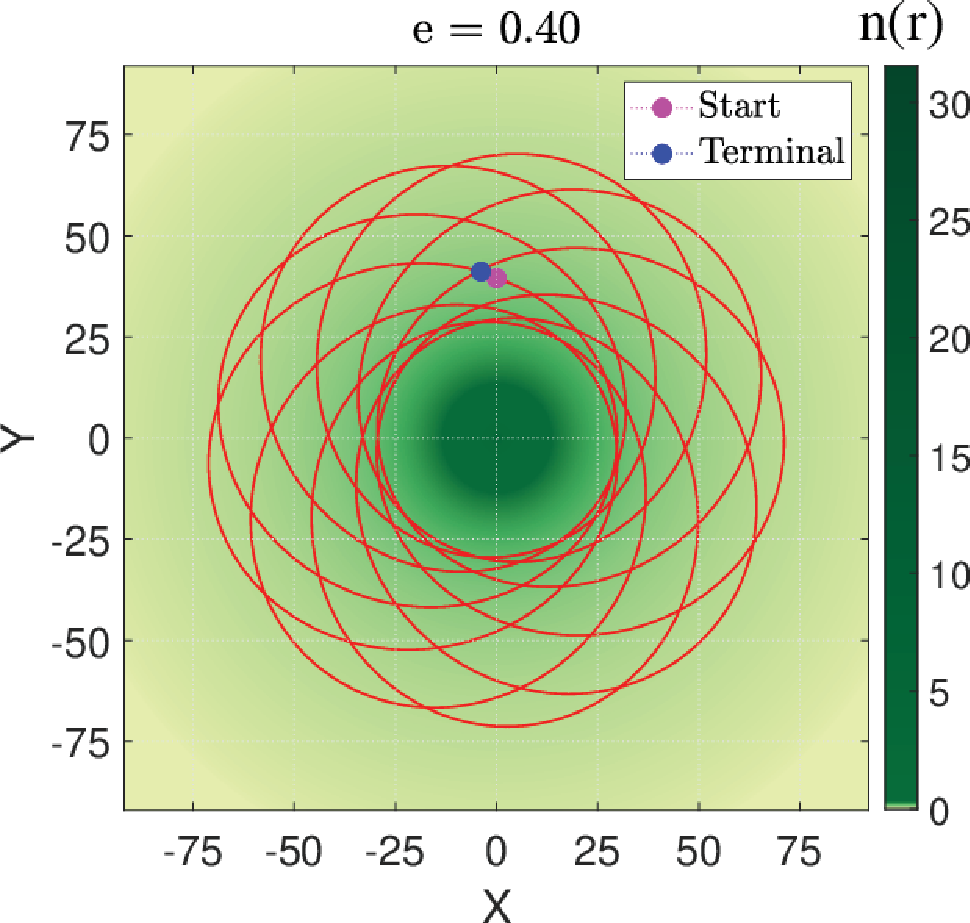}
\label{fig:precess_orbits_c}}
\quad
\subfigure[Orbit in precession with $e=0.8$.]{
\includegraphics[width=0.35\textwidth]{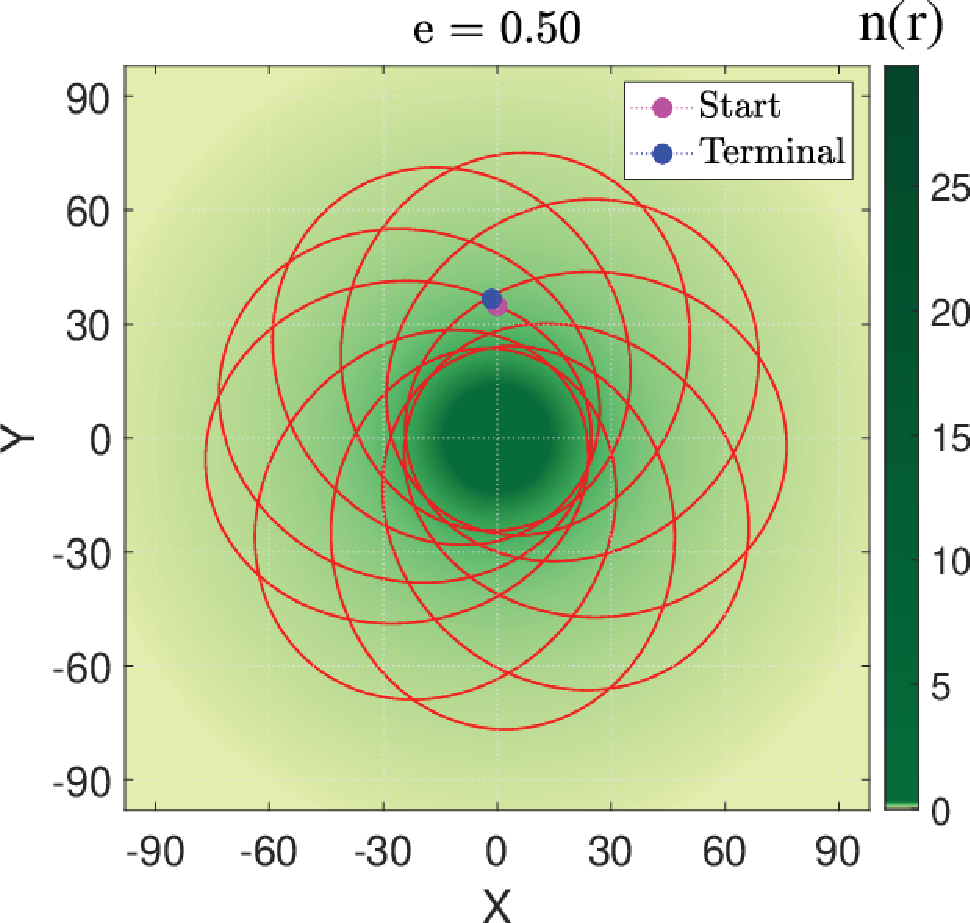}
\label{fig:precess_orbits_d}}
\quad
\caption{The trajectories of light rays in the Schwarzschild-like medium with the same $\alpha=0.1$ and 
different $e$ from $0.2$ to $0.5$ respectively. Each light ray follows a closed precessing orbit, since $\alpha=0.1$ is a rational number. 
The larger $e$ is, the further away the perihelion and aphelion are seperated from each other.}
\label{fig:precess_orbits}
\end{figure}

For light to follow a similar orbit with pericenter precession, all we need to do is to customize the index profile $n(r)$ 
by simply relating it to $v(r)$ as we did in eq.\ \eqref{eqn:refractive_index1}. 
This basic relation arises from the analogy between classical mechanics and optics that in both areas 
the least action principle plays a center role in determining the moving path of any object, 
whether it is a mechanical particle or a massless photon. In this way, $n(r)$ can be designated as: 
\begin{equation}\label{eqn:refractive_index2}
n(r)=n_0 \left\{ \frac{1}{r}-\frac{1}{2a(2E-1)}+\left[ 2GM+\frac{p}{2(2E-1)}\cdot \frac{\alpha(2-\alpha)}{(1-\alpha)^2} \right] \frac{1}{r^2} \right\}^{\frac{1}{2}}
\end{equation}

where $p$ and $a$ are structural parameters of the quasi-elliptical orbit, so we can use them to cumtomize the scale and shape of light orbits. 
The value of $\alpha$ (usually $\alpha\ll 1$) reflects the extent to which relativistic effects emerge to have an apparent impact. 
Once $n(r)$ is specified, the trajectories of light rays will follow the E-L equation given in eq.\ \eqref{eqn:Binet equation}, 
with the initial conditions $r(\phi_i)$ and $r^\prime(\phi_i)$: 
\begin{equation}\label{eqn:initial_cond2}
\begin{split}
&\left\{
\renewcommand\arraystretch{2}
\begin{array}{ll}
r(\phi_i) & =\dfrac{p}{1-e\cos(1-\alpha)(\phi_i-\phi_0)} \\
r^\prime(\phi_i) & =-\dfrac{pe(1-\alpha) \sin(1-\alpha)(\phi_i-\phi_0)}{\left[ 1-e\cos(1-\alpha)(\phi_i-\phi_0) \right]^2} \\
\ & =-\dfrac{r^2(\phi_i) e(1-\alpha)}{p} \sin(1-\alpha)(\phi_i-\phi_0)
\end{array}\right.
\\
\vspace{1.5em}
&\Rightarrow\ 
\left\{
\renewcommand\arraystretch{2}
\begin{array}{ll}
p & =\dfrac{(1-\alpha)^2 \left[ 2-\dfrac{r(\phi_i)}{a} \right] r^3(\phi_i)}
{(1-\alpha)^2 r^2(\phi_i)+r^{\prime 2}(\phi_i)}  \\
\phi_0 & =\phi_i-\dfrac{1}{1-\alpha} \arctan \dfrac{r^\prime(\phi_i) p}{r(\phi_i) (1-\alpha) \left[ p-r(\phi_i) \right]} \\
\ & =\phi_i-\dfrac{1}{1-\alpha} \arctan \dfrac{(1-\alpha) \left[ 2-\dfrac{r(\phi_i)}{a} \right] r(\phi_i) r^\prime(\phi_i)}
{(1-\alpha)^2 \left[ 1-\dfrac{r(\phi_i)}{a} \right] r^2(\phi_i)-r^{\prime 2}(\phi_i)}
\end{array}\right.
\end{split}
\end{equation}

from which the integration constants $p$ and $\phi_0$ are determined, and then the eccentricity $e$. 
The E-L equation with $n(r)$ in eq.\ \eqref{eqn:refractive_index2} can be solved numerically. 
As an example, we set $\alpha=0.1$ and choose appropriate $r(\phi_i)$ and $r^\prime(\phi_i)$ to get the values of $e=0.2,\ 0.3,\ 0.4,\ 0.5$ respectively, 
and the simulation results are shown in Fig.\ \ref{fig:precess_orbits}. 

As Fig.\ \ref{fig:precess_orbits} shows, the geodesic light follows with the shortest optical path is a quasi-elliptical orbit in precession, 
of which the pericentric point advances by $2\alpha\pi/(1-\alpha)=40^o$ when light finishes one cycle of evolution. 
After $9$ cycles, light will return to the initial point, which forms a closed petal-like orbit. 
The result reproduces well the pericenter precession of a planet revolving around a massive star. 
Furthermore, we notice that the general condition for a closed orbit is determined by the value of $\alpha$ defined before 
eq.\ \eqref{eqn:orbit_chi}. If $\alpha(1-\alpha)^{-1}$ is a rational number, it can be written as: 
\begin{equation}
\frac{\alpha}{1-\alpha}=\left\lfloor \frac{\alpha}{1-\alpha} \right\rfloor+\frac{u}{w}
\end{equation}

where $\lfloor \cdot \rfloor$ means taking the integer part of input, and $u/w$ is a irreducible fraction such that $u$ and $w$ are co-primes. 
Then the orbit is closed because light will return to the initial point after traveling for $\lfloor \cdot \rfloor*w+u$ cycles around the center of the medium, 
while passing through the perihelions for $w$ times. 

Light imitates planetary motion in relativistic limit in the Schwarzschild medium. The optical properties brought by this analogy can be further revealed 
by examing the orbits of multiple light rays in the medium. To model multiple light rays, we adjust $\alpha$ and $\phi_0$ simultaneously 
in eq.\ \eqref{eqn:initial_cond2}, to keep either $r(\phi_i)$ or $r^\prime(\phi_i)$ constant and change only one of them. If $r^\prime(\phi_i)$ is fixed and $r(0)$ varies, 
it corresponds to a bunch of parallel rays, whereas if only $r^\prime(\phi_i)$ varies, it corresponds to light rays emitting from a point source toward different directions. 

\begin{figure}[htp]
\centering
\subfigure[Parallelly incident light rays.]{
\includegraphics[width=0.4\textwidth]{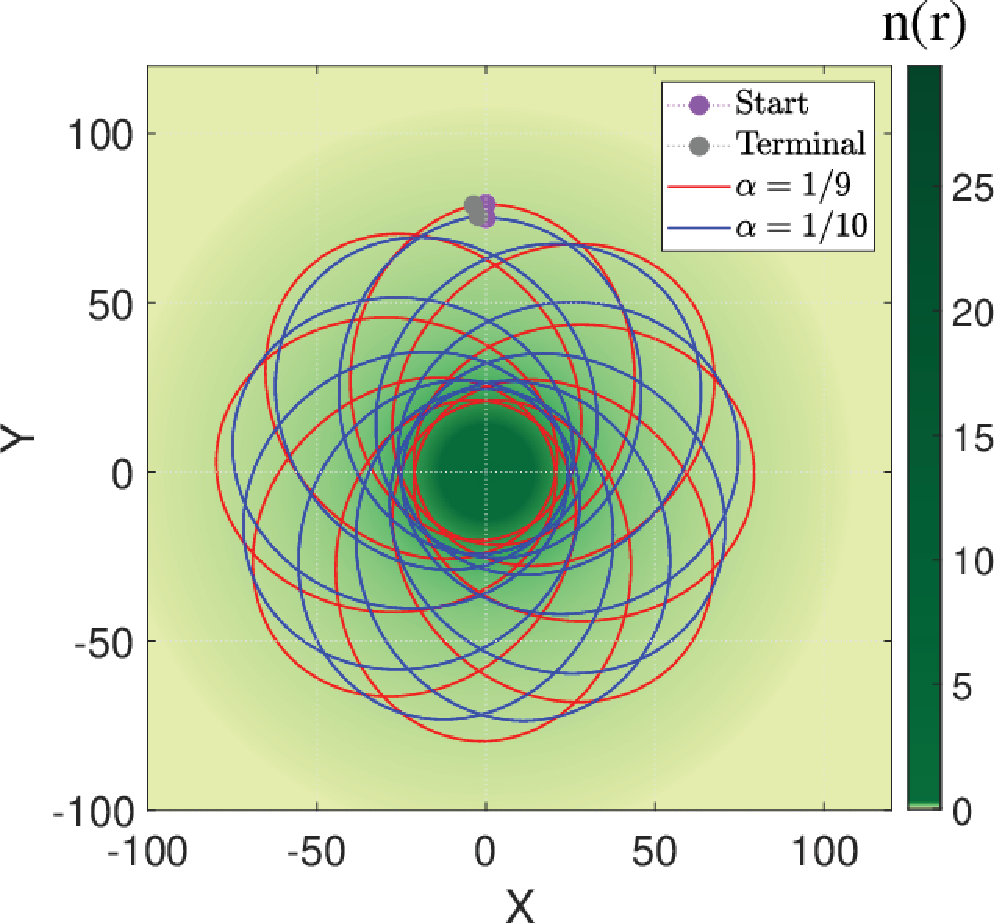}
\label{fig:multi_precess_orbits_a}}
\quad
\subfigure[Light rays from a point source.]{
\includegraphics[width=0.4\textwidth]{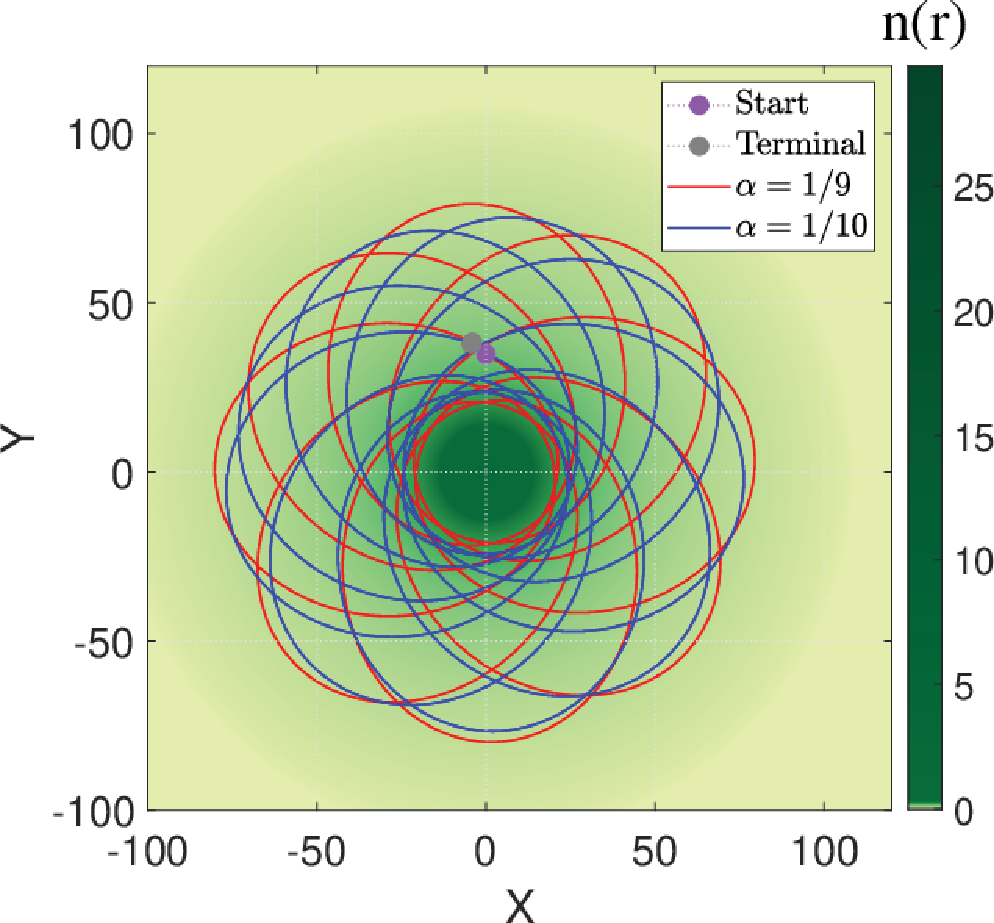}
\label{fig:multi_precess_orbits_b}}
\\
\caption{(a) Light emits parallely into the Schwarzschild medium. 
These light rays have the same incident direction but different orbital parameters that satisfy different initial conditions. 
We depict two rays with $\alpha=0.1,\ 0.2$; what they follow are distinct trajectories marked by different precession angles. 
However, they finally return to the initial point since both orbits are closed. 
(b) Light emitting from a point source along different directions into the Schwarzschild medium. The trajectories of two rays with $\alpha=0.1,\ 0.2$ are shown. 
Similar arguments can be made on these rays that they undergo different orbits and converge at the same initial point.}
\label{fig:multi_precess_orbits}
\end{figure}

As Fig.\ \ref{fig:multi_precess_orbits} shows, each light ray has distinct $\alpha$ and $e$, 
orbiting around the refractive index center in paths of different shapes and precession angles. 
For both Fig.\ \ref{fig:multi_precess_orbits_a} and \ref{fig:multi_precess_orbits_b}, $\alpha$ takes rational values, 
so the trajectories form closed orbits. 

In this sense, light exhibits complicated behaviors that are sensitive to the initial conditions. 
The various closed orbits imply the complexity of resonance phenomena in the Schwarzschild medium, which can be viewed as a multipath optical cavity. 
In particular, if $\alpha$ is small, light needs to complete so many cycles before returning to the start point, 
that the effective optical path can be extremely long. Taking advantage of this fact, 
we can manipulate the resonance properties of the Schwarzschild medium by tuning the value of $\alpha$. 

\subsection{Period of non-elliptical light orbits}
The next question we care about is whether the precession of perihelion has influence on the period of light orbits. 
We start from eq.\ \eqref{eqn:period_integ} again, and rewrite the integral by using the relation $|d\mathbi{r}|=v(\mathbi{r}) d\tau$ as: 
\begin{equation}
\begin{split}
T&= \frac{1}{c}\oint n(\mathbi{r}) |d\mathbi{r}|
=\frac{1}{c} \int_{\tau(0)}^{\tau(T)} n(r)v(r) d\tau \\
&= \frac{1}{cJ} \int_{0}^{\frac{2\pi}{1-\alpha}} n(r)v(r)r^2 d\phi
\end{split}
\end{equation}

Because $n(r)\propto v(r)$, the above equation implies that $T\propto\int n^2(r)r^2 d\phi$, 
which resembles the optical Binet equation for ray tracing in spherical GRIN media \cite{Binet}. 
Therefore we have: 
\begin{equation}
\begin{split}
T&= \frac{(1-\alpha)\sqrt{2GM(2E-1)}}{cn_0 J} \int_{0}^{\frac{2\pi}{1-\alpha}} n^2(r)r^2 d\phi \\
&= \frac{(1-\alpha) n_0 \sqrt{2GM}}{cJ}
\int_{0}^{\frac{2\pi}{1-\alpha}} \left[ (r+2GM)(2E-1)-\frac{r^2}{2a}+\frac{\alpha(2-\alpha) p}{2(1-\alpha)^2} \right] d\phi
\end{split}
\end{equation}

Note that the integration is evaluated from $0$ to $2\pi(1-\alpha)^{-1}$, to take into account the advancement of the pericenter during one cycle of the quasi-circular orbit. 
We make a change of variables by $\psi=(1-\alpha)\phi$ and substitute $r$ and $r^2$ in the integrand by eq.\ \eqref{eqn:orbit_precess}, 
the integration can be carried out analytically based on the results in eq.\ \eqref{eqn:integral_r} and \eqref{eqn:integral_r2}: 
\begin{equation}
\begin{split}
T&= \frac{n_0 \sqrt{2GM}}{cJ}
\int_{0}^{2\pi} \left[ (r+2GM)(2E-1)-\frac{r^2}{2a}+\frac{\alpha(2-\alpha) p}{2(1-\alpha)^2} \right] d\psi \\
&= \frac{2\pi n_0}{c} \sqrt{\frac{p}{2}} \left[ \frac{4E-3}{\sqrt{1-e^2}}+\frac{4GM(2E-1)}{p}+\frac{\alpha(2-\alpha)}{(1-\alpha)^2} \right] \\
&= \frac{2\pi n_0}{c} \left\{ (4E-3)\sqrt{\frac{a}{2}}+\left[ \frac{4\alpha}{3} (2E-1)+\frac{\alpha(2-\alpha)}{(1-\alpha)^2} \right] \sqrt{\frac{p}{2}} \right\}
\end{split}
\end{equation}

On the other hand, the number of cycles that light trajectories go through, denoted by $m$, 
for the pericenter to advance by $2\pi$ is $m=2\pi/\Delta\psi=(1-\alpha)/\alpha$, 
so the amount of time elapsed during this process, i.e.\ the full period of a closed light orbit, is: 
\begin{equation}\label{eqn:accum_period}
\begin{split}
mT&= 
\frac{2\pi n_0}{c} \left\{ \frac{(1-\alpha)(4E-3)}{\alpha} \sqrt{\frac{a}{2}}+\left[ \frac{4(1-\alpha)(2E-1)}{3}+\frac{2-\alpha}{1-\alpha} \right] \sqrt{\frac{p}{2}} \right\} \\
&\approx mT_0+\Delta T \\
\text{where}\ \Delta T&= \frac{2\pi n_0}{c}
\left[ \frac{4(1-\alpha)}{3}+\frac{2-\alpha}{1-\alpha} \right] \sqrt{\frac{p}{2}}\approx \frac{2\pi n_0}{c} \cdot\frac{10}{3} \sqrt{\frac{\,p\,}{2}}
\end{split}
\end{equation}

\begin{figure}[htp]
\centering
\includegraphics[width=0.45\textwidth]{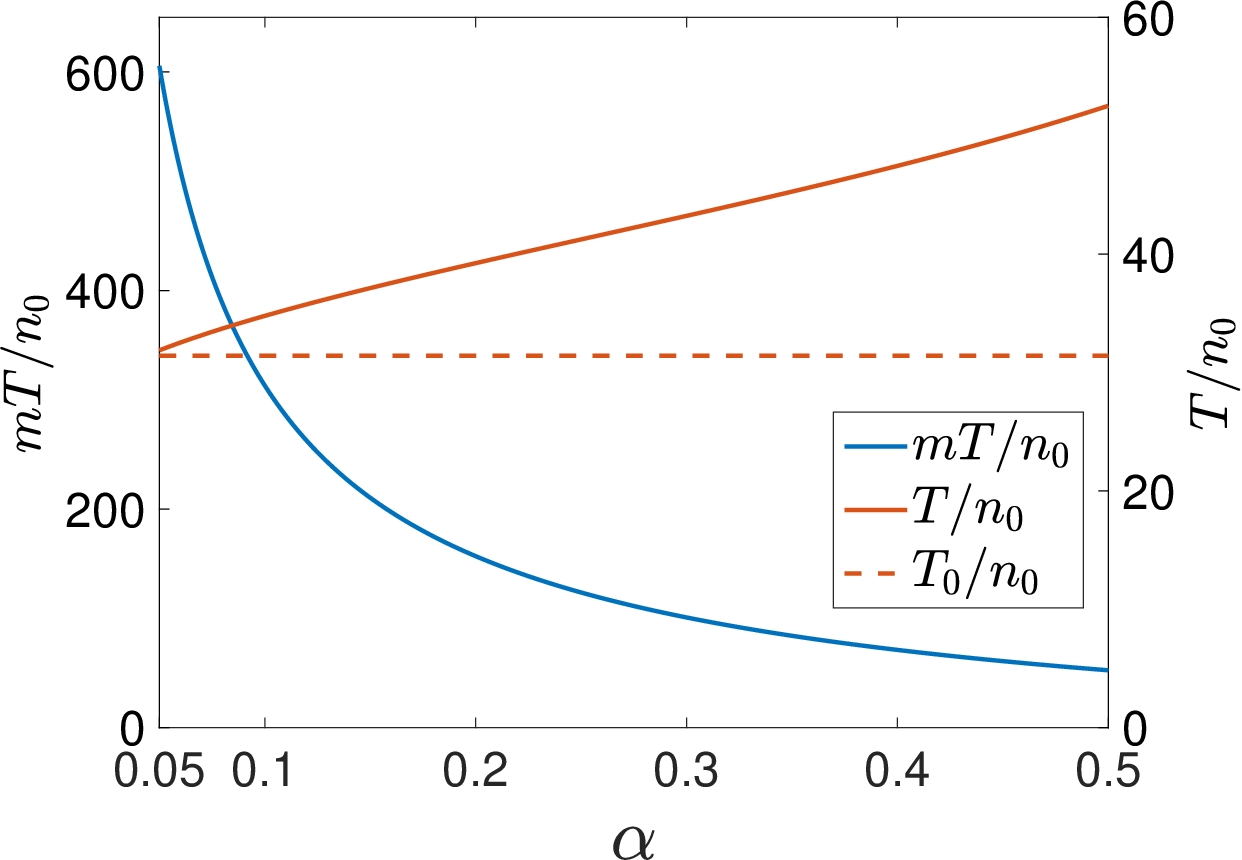}
\caption{Left axis: period $T$ of one cycle of the precessing orbit, the coordinate time elapsed on the path in between two neighboring pericenters. 
Right axis: period $mT$ of multiple cycles that form a closed orbit, on which the pericenter shift accumulates to integer multiples of $2\pi$.
The red dash line indicates the period $T_0$ of elliptical orbits discussed in section \ref{sec:ellipse_period} as a reference.}
\label{fig:period}
\end{figure}

Here $T_0$ is from eq.\ \eqref{eqn:period_Newton}, representing the period of an elliptical orbit with the same $a$ in Newtonian gravity. 
The first term in eq.\ \eqref{eqn:accum_period} depends on $m$, which is then dependent on $\alpha$, so it is different from one ray to another. 
The second term $\Delta T\propto \sqrt{p/2}$ and therefore is specific to each different light ray as well. 
It represents the relativistic correction accumulated in $m$ cycles. 
The variation of $T$ and $mT$ as a function of $\alpha$ is shown in Fig.\ \ref{fig:period}. 

It is found that $T$ increases with $\alpha$, mainly due to the fact that a larger $\alpha$ results in a larger increment of $\phi$ 
and hence a longer path, during a cycle. 
On the other hand, $mT$ decays since there are less cycles needed to complete a closed orbit as $\alpha$ increases. 
This information is useful for computing and comparing the time delay of light propagating in complex orbits in a general radially GRIN medium. 
It is worth noticing that, the precession angle will exceed $2\pi$ towards a larger $\alpha$ than $0.5$, 
such that the orbits will gradually become more eccentric and a variety of orbital patterns will emerge in this process \cite{periodicity}. 

We also draw $T_0$, the period of elliptical orbits under Newton's gravity, as a baseline in Fig.\ \ref{fig:period} in a red dotted line. 
One can clearly see that when $\alpha$ decreases down to $0$, $T$ tends to be identical to $T_0$. In this process, 
because the precessing orbit gradually becomes an elliptical orbit, the variation of $T$ is consistent with the change of orbit. 

\section{Geometric model}
\subsection{Optical medium analogous to Schwarzschild spacetime}
When light travels in space, it does not move along planetary orbits in gravitational fields, since there is a fundamental difference 
between light and planets. Light in nature is a bunch of massless photons, while a planet consists of massive particles. 
No four velocity can be defined for light. However, the geodesic equation that light follows 
can be derived in terms of four momentum $\mathbi{p}$ from parallel transport of $\mathbi{p}$ along a geodesic: 
\begin{equation}
\frac{dp^\rho}{d\lambda}+\Gamma_{\mu\nu}^{\rho} p^\mu p^\nu=0
\end{equation}

where $\lambda$ is an affine parameter. The $r$-component of this equation reads: 
\begin{equation}\label{eqn:geodesics4}
\frac{d^2 r}{d\lambda^2}=\frac{GM}{r^2} \left( 1-\frac{2GM}{r} \right)^{-1} \left[ E^2-\left( \frac{dr}{d\lambda} \right)^2 \right]
-\left( 1-\frac{2GM}{r} \right) \frac{J^2}{r^3}
\end{equation}

Note that the contancy of $p_t=-E$ and $p_\phi=J$ have been used in deriving eq.\ \eqref{eqn:geodesics4}. 
In the meantime, the fact that a light path is a null line requires $\mathbi{p} \cdot \mathbi{p}=0$, which can be written in components: 
\begin{equation}\label{eqn:zero_momentum}
-\left( 1-\frac{2GM}{r} \right)^{-1} E^2+\left( 1-\frac{2GM}{r} \right)^{-1} \left( \frac{dr}{d\lambda} \right)^2+\frac{J^2}{r^2}=0
\end{equation}

Combining eqs.\ \eqref{eqn:geodesics4} and \eqref{eqn:zero_momentum}, we arrive at the following result: 
\begin{equation}
\begin{split}
\frac{d^2 r}{d\lambda^2}&= \frac{GM}{r^2} \frac{J^2}{r^2}-\left( 1-\frac{2GM}{r} \right) \frac{J^2}{r^3} \\
&= -\frac{J^2}{r^3}\left( 1-\frac{3GM}{r} \right)
\end{split}
\end{equation}

Following similar procedures with solving the euqation of motion for a massive particle, 
we introduce $u=r^{-1}$ and change the variable of differentiation from $\lambda$ to $\phi$ by using the relation: 
\begin{equation}
\frac{d^2r}{d\lambda^2}=-J^2u^2 \frac{d^2u}{d\phi^2}
\end{equation}

Then we find the geodesic equation for light: 
\begin{equation}\label{eqn:geodesics3_u}
\frac{d^2u}{d\phi^2}+u=3GMu^2.
\end{equation}

Treating the $u$-squared term on the right side as a pertubation, this equation can be solved to give: 
\begin{equation}\label{eqn:light_orbit2}
u=d\cos(\phi-\phi_0)+GMd^2 \left[ 2-\cos^2(\phi-\phi_0)  \right]
\end{equation}

In the solution $d$ and $\phi_0$ are integral constants, which can be determined from the initial conditions: 
\begin{equation}\label{eqn:initial_condition3}
\left\{
\renewcommand\arraystretch{2}
\begin{array}{ll}
r(\phi_i)&= \dfrac{1}{d\cos(\phi_i-\phi_0)+GMd^2 \left[ 2-\cos^2(\phi_i-\phi_0)  \right]} \\
r^\prime(\phi_i)&= \dfrac{d\sin\left( \phi_i-\phi_0 \right)-GMd^2 \sin2 \left( \phi_i-\phi_0 \right)}
{\left\{ d\cos(\phi_i-\phi_0)+GMd^2 \left[ 2-\cos^2(\phi_i-\phi_0)  \right] \right\}^2} \\
\ &= r^2(\phi_i) d\sin\left( \phi_i-\phi_0 \right) \left[ 1-2GMd \cos\left( \phi_i-\phi_0 \right) \right]
\end{array}\right.
\end{equation}

Inserting the first equation into the second one gives: 
\begin{equation}\label{eqn:d_phi0}
\begin{split}
\frac{r^\prime(\phi_i)}{\tan\left( \phi_i-\phi_0 \right)}&= r(\phi_i)-GMd^2 r^2(\phi_i) \cdot
\frac{2\tan^2\left( \phi_i-\phi_0 \right)+3}{\tan^2\left( \phi_i-\phi_0 \right)+1} \\
\Rightarrow\ GMd^2&= \frac{1}{r(\phi_i)} \cdot \frac{\tan^2\left( \phi_i-\phi_0 \right)+1}{2\tan^2\left( \phi_i-\phi_0 \right)+3}
-\frac{r^\prime(\phi_i)}{r^2(\phi_i)} \cdot \frac{\tan\left( \phi_i-\phi_0 \right)+\cot\left( \phi_i-\phi_0 \right)}{2\tan^2\left( \phi_i-\phi_0 \right)+3}
\end{split}
\end{equation}

This result enables us to eliminate the variable $d$ in eq.\ \eqref{eqn:initial_condition3} and solve for $\phi_0$. 
Alternatively, setting eq.\ \eqref{eqn:light_orbit2} equal to $0$ gives the incident direction $\phi=\phi_0+\pi-\arccos(2GMd)\approx \phi_0+\pi/2$ 
at $r=+\infty$. For light incident along a particular direction, 
$\phi_0$ can be deduced from this relation, and thereforth $d$ can be determined from eq.\ \eqref{eqn:d_phi0}. 

\begin{figure}[htp]
\centering
\subfigure[Light path with $\Delta\phi=1.154\pi$.]{
\includegraphics[width=0.4\textwidth]{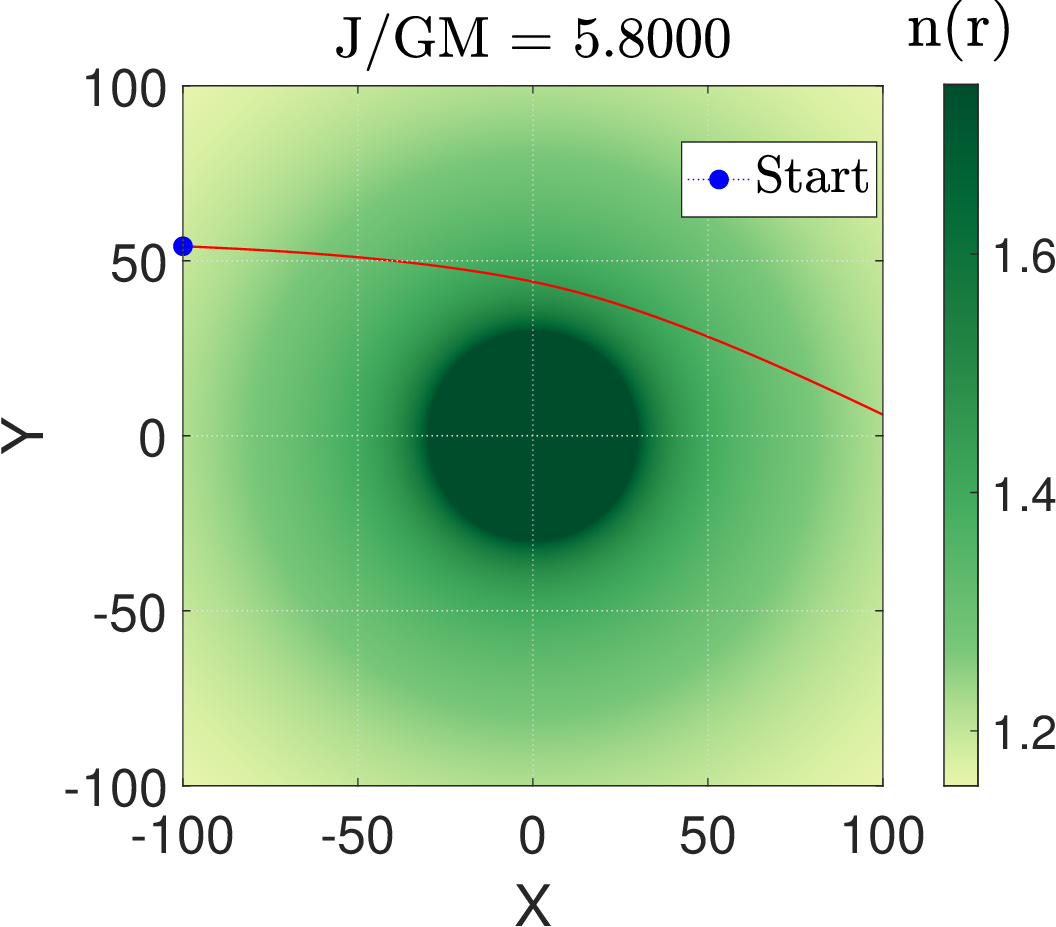}
\label{fig:deflect_orbits_a}}
\quad
\subfigure[Light path with $\Delta\phi=1.475\pi$.]{
\includegraphics[width=0.4\textwidth]{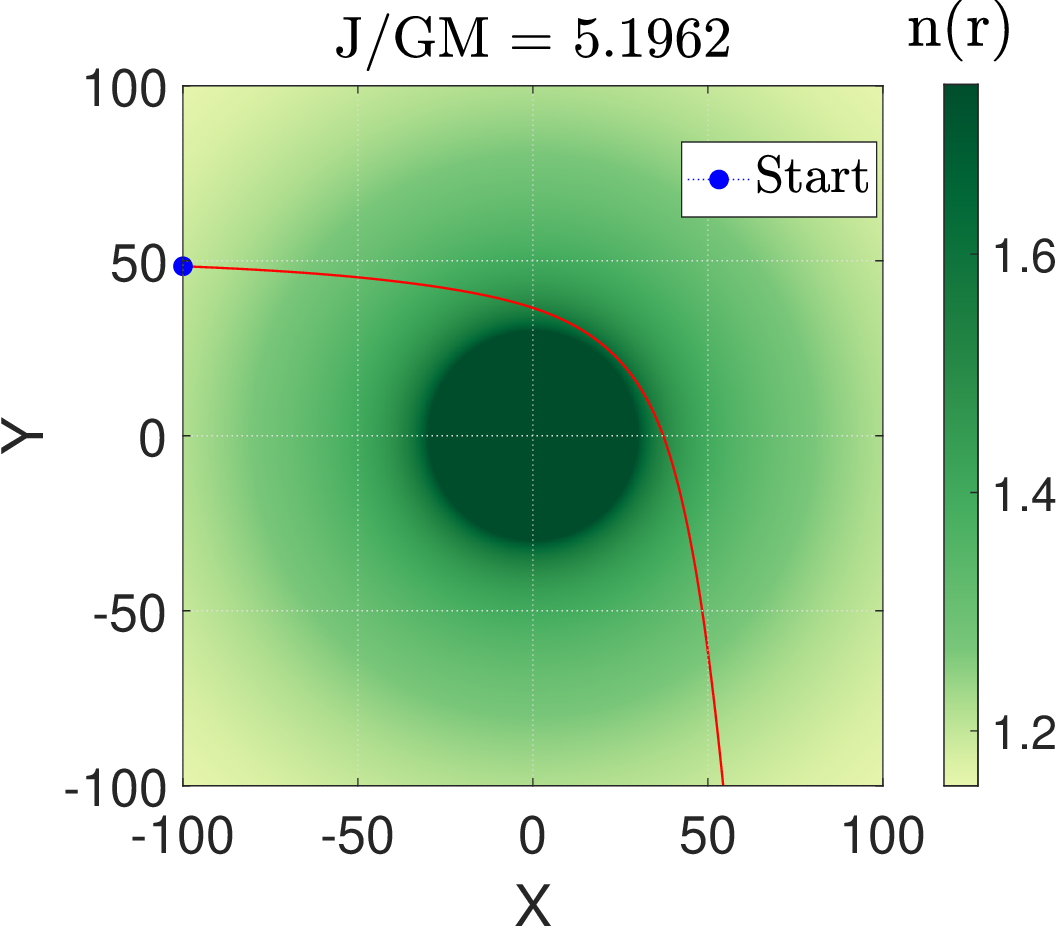}
\label{fig:deflect_orbits_b}}
\\
\subfigure[Light path with $\Delta\phi=2.019\pi$.]{
\includegraphics[width=0.4\textwidth]{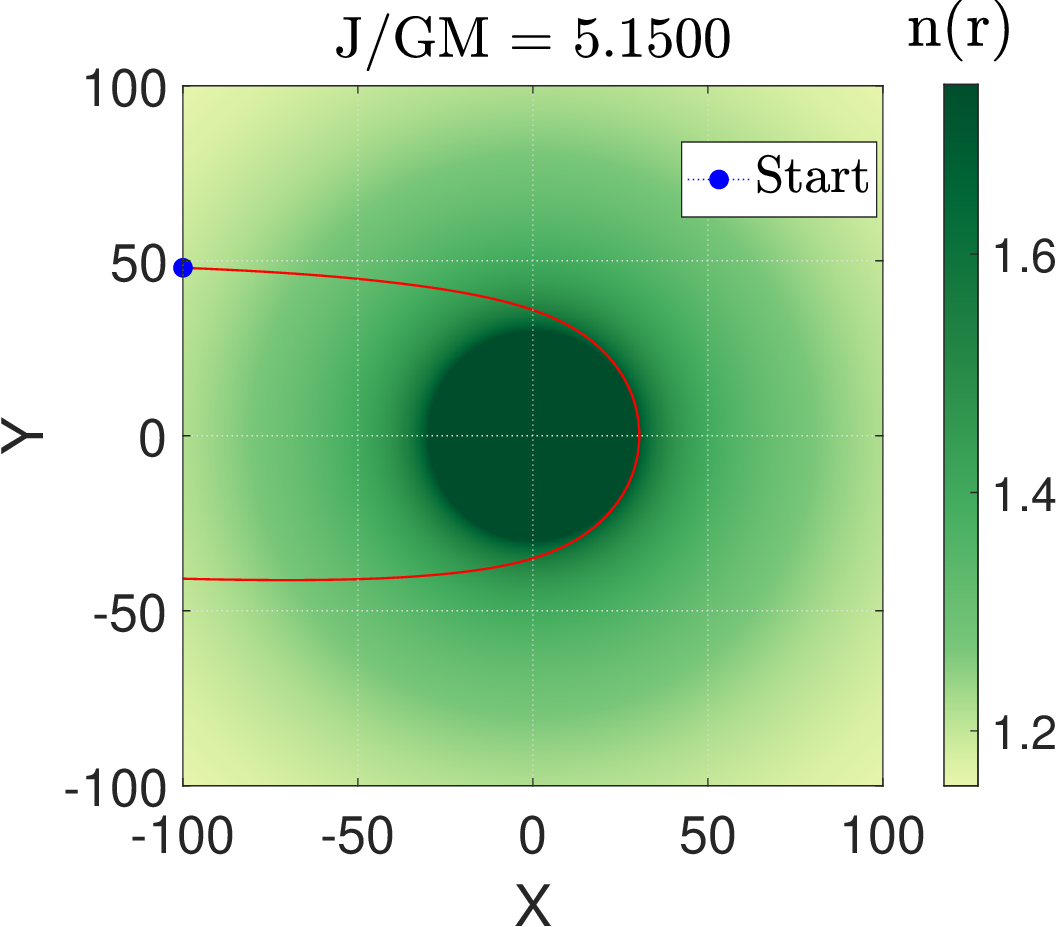}
\label{fig:deflect_orbits_c}}
\quad
\subfigure[Light path with $\Delta\phi=2.793\pi$.]{
\includegraphics[width=0.4\textwidth]{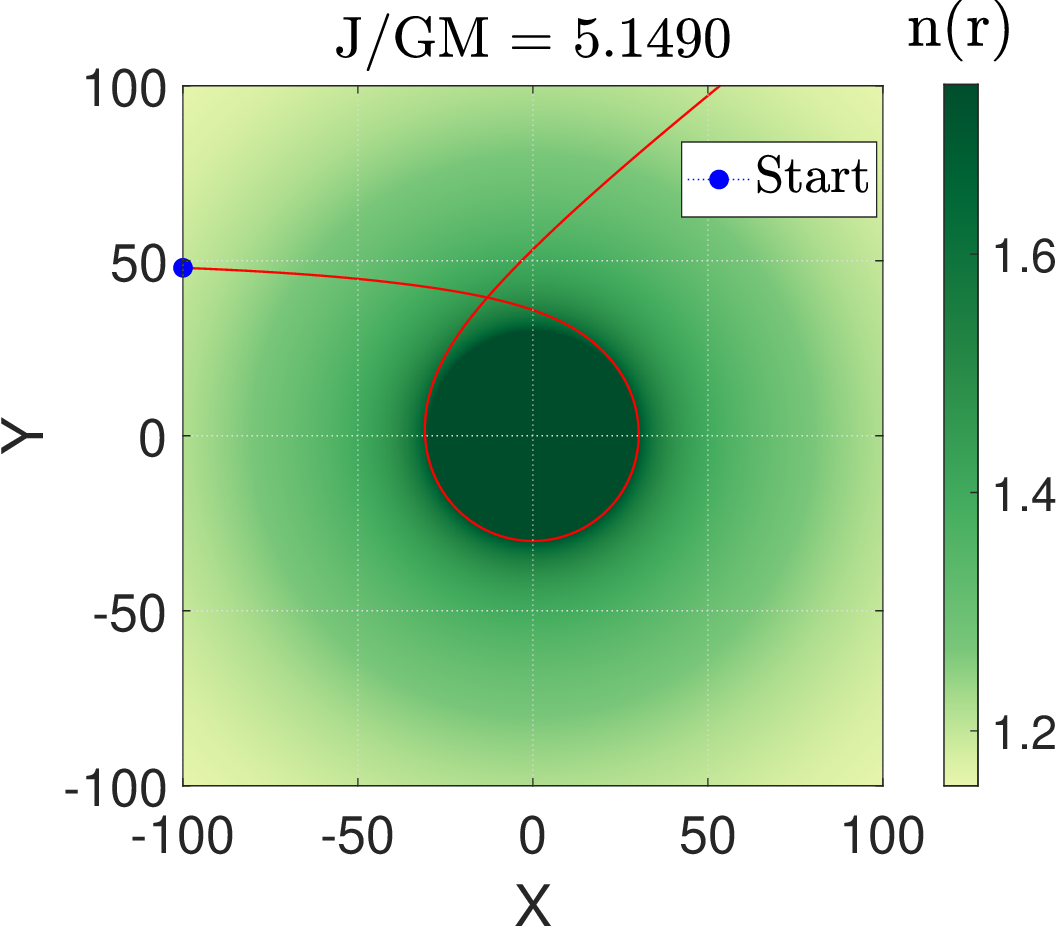}
\label{fig:deflect_orbits_d}}
\\
\caption{Light orbits with different angular momentum $J$ in index profile imitating gravitational fields. 
When $J$ approaches $J_m$, light tends to stay on the critical circular orbit, close to which $\Delta\phi$ increases rapidly.}
\label{fig:deflect_orbits}
\end{figure}

Imagine that we are observing light trajectories in a flat space filled with some optical medium, 
and we attribute the bending of light to some refractive index profile $n(r,\phi)$ in the medium. 
The apparent velocity of light measured in observer's coordinates $(t,r,\phi)$ with the metric in eq.\ \eqref{eqn:Schwarzschild} appears to be: 
\begin{equation}
\begin{split}
v^2(r,\phi)&= \left( \frac{dr}{dt} \right)^2+r^2 \left( \frac{d\phi}{dt} \right)^2 \\
&= -\frac{g_{tt}}{g_{rr}} \left[ 1-\frac{p_\phi^2}{g_{rr} g_{\phi\phi}} \left( 1-\frac{g_{rr} r^2}{g_{\phi\phi}} \right) \right]
\end{split}
\end{equation}

The last line results from the fact that $t$ is related to the generic affine parameter $\lambda$ by $\lambda=\sqrt{\left| g_{tt}/g_{rr} \right|} t+\lambda_0$, 
where $\lambda_0$ is a constant. It follows that: 
\begin{equation}
\frac{d\phi}{dt}=\sqrt{\frac{\left| g_{tt} \right|}{g_{rr}}} \frac{d\phi}{d\lambda}=\sqrt{\frac{\left| g_{tt} \right|}{g_{rr}}} \frac{p_\phi}{g_{\phi\phi}}
\end{equation}

Without losing generality, we can choose $\lambda_0=0$ so that $\lambda$ is proportional to $t$ and rescaled by the metric tensor. 
On the other hand, an observer may attribute the spacetime geometry to the existence of an optical medium with a certain index profile $n(r,\phi)$, 
by relating the velocity profile $v(r,\phi)$ to $n(r,\phi)$ according to $v(r,\phi)=n^{-1}(r,\phi)$. 
Consequently, viewed from the observer's perspective, the behavior of light subject to spacetime geometry 
can be described equivalently by ray optics in a dielectric environment with the refractive index: 
\begin{equation}
n^2(r,\phi)
=-\frac{\dfrac{g_{rr}}{g_{tt}}}{1-\dfrac{p_\phi^2}{g_{rr} g_{\phi\phi}} \left( 1-\dfrac{g_{rr} r^2}{g_{\phi\phi}} \right)}
\end{equation}

For Schwarzschild metric, the expression for $n(r,\phi)$ is: 
\begin{equation}\label{eqn:refractive_index3}
n(r,\phi)=\frac{1}{\left( 1-\dfrac{2GM}{r} \right) \sqrt{1+\dfrac{2GMJ^2}{r^3}}}
\end{equation}

In this case, $n(r,\phi)$ is a function of $r$ only, as it should be in a spherically symmetric geometry. 
We then appoint $n(r)$ in the E-L equation equal to eq.\ \eqref{eqn:refractive_index3} to simulate the orbits of light, 
and the results are shown in Fig.\ \ref{fig:deflect_orbits}. These orbits agree well with light paths in gravitational fields \cite{gravitation_lensing}; 
they are the analogues of gravitational deflection and lensing of light in astronomy. 

\subsection{Deflection angle of light orbits}
The orbital equation of light can be derived from the components of photons' four momentum $\mathbi{p}$. 
While the covariant components $p_t=E$ and $p_\phi=J$ are conserved quantities, the radial component $dr/d\lambda$ follows 
from the identity $\mathbi{p}\cdot \mathbi{p}=0$ that: 
\begin{equation}\label{eqn:radial_comp}
\left( \frac{dr}{d\lambda} \right)^2=-\frac{1}{g_{rr}}\left( \frac{p_t^2}{g_{tt}}+\frac{p_\phi^2}{g_{\phi\phi}} \right)
\end{equation}

In the meantime, $d\phi/d\lambda=p_\phi/g_{\phi\phi}$, so we obtain: 
\begin{equation}\label{eqn:angle_element1}
\left( \frac{d\phi}{dr} \right)^2=-\frac{g_{rr}}{g_{\phi\phi}} \frac{g_{tt} p_\phi^2}{g_{\phi\phi} p_t^2+g_{tt} p_\phi^2}
=\frac{1}{r^2\left[ \dfrac{r^2}{b^2}-\left( 1-\dfrac{2GM}{r} \right) \right]}
\end{equation}

where $b$ is known as the impact parameter in scattering theory, defined as $b=J/E$, 
since the deflection of light in gravitational field is similar to the scattering problem of mechanical particles in a central force field. 
For a particular light ray, the geometrical meaning of $b$ is the distance from the incident direction to the center of gravitational attraction \cite{Leonhardt}. 
In our units, $E\approx 1$ such that $b\approx J$. 
Because $\mathbi{g}$ is independent of $\phi$, light trajectories with $\phi$ going clockwise and counter-clockwise are just mirror images of each other. 
Therefore we only need to consider one of the cases, say counter-clockwise, in which $d\phi/dr>0$, 
and take the positive square root of eq.\ \eqref{eqn:angle_element2}: 
\begin{equation}\label{eqn:angle_element2}
\frac{d\phi}{dr}=\frac{1}{r \sqrt{\dfrac{r^2}{J^2}-\left( 1-\dfrac{2GM}{r} \right)}}
\end{equation}

\begin{figure}[htp]
\subfigure[Pericentric distance $r_m$ versus $J/GM$.]{
\includegraphics[width=0.4\textwidth]{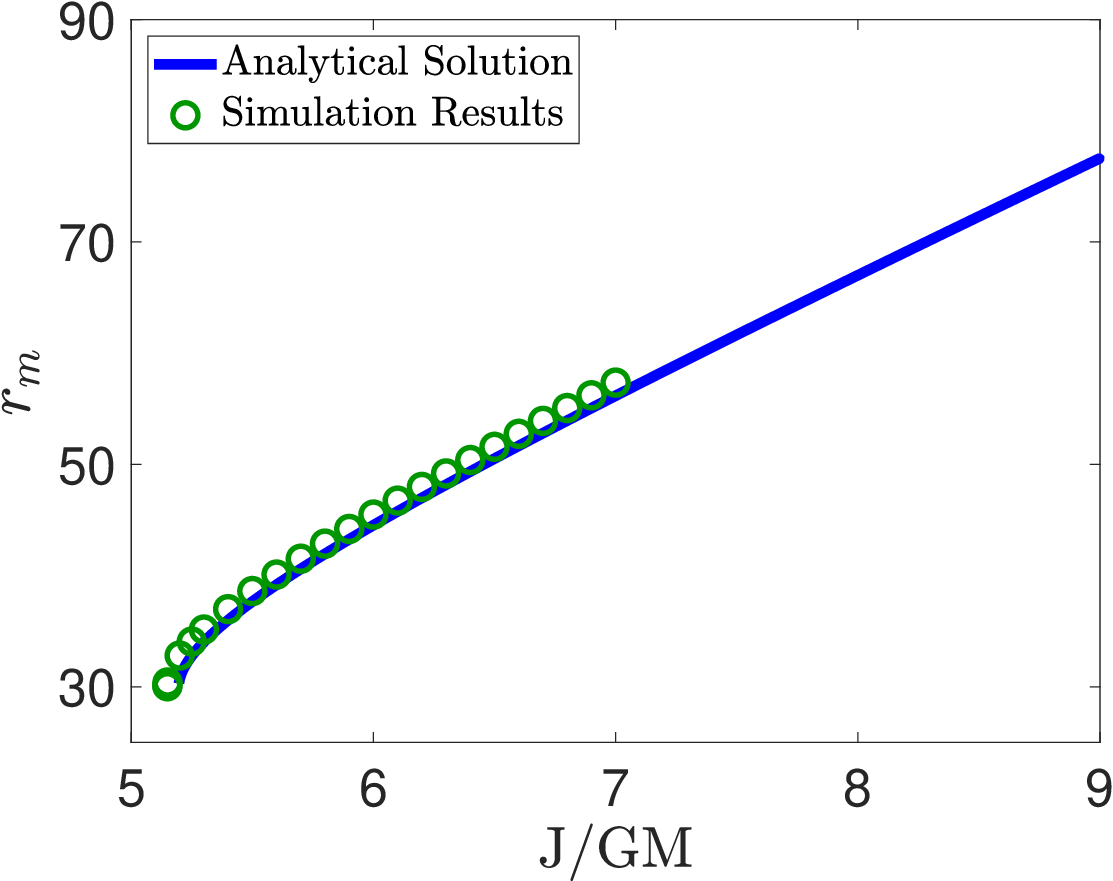}
\label{fig:deflect_angle_a}}
\quad
\subfigure[Deflection angle $\Delta\phi$ versus $J/GM$.]{
\includegraphics[width=0.408\textwidth]{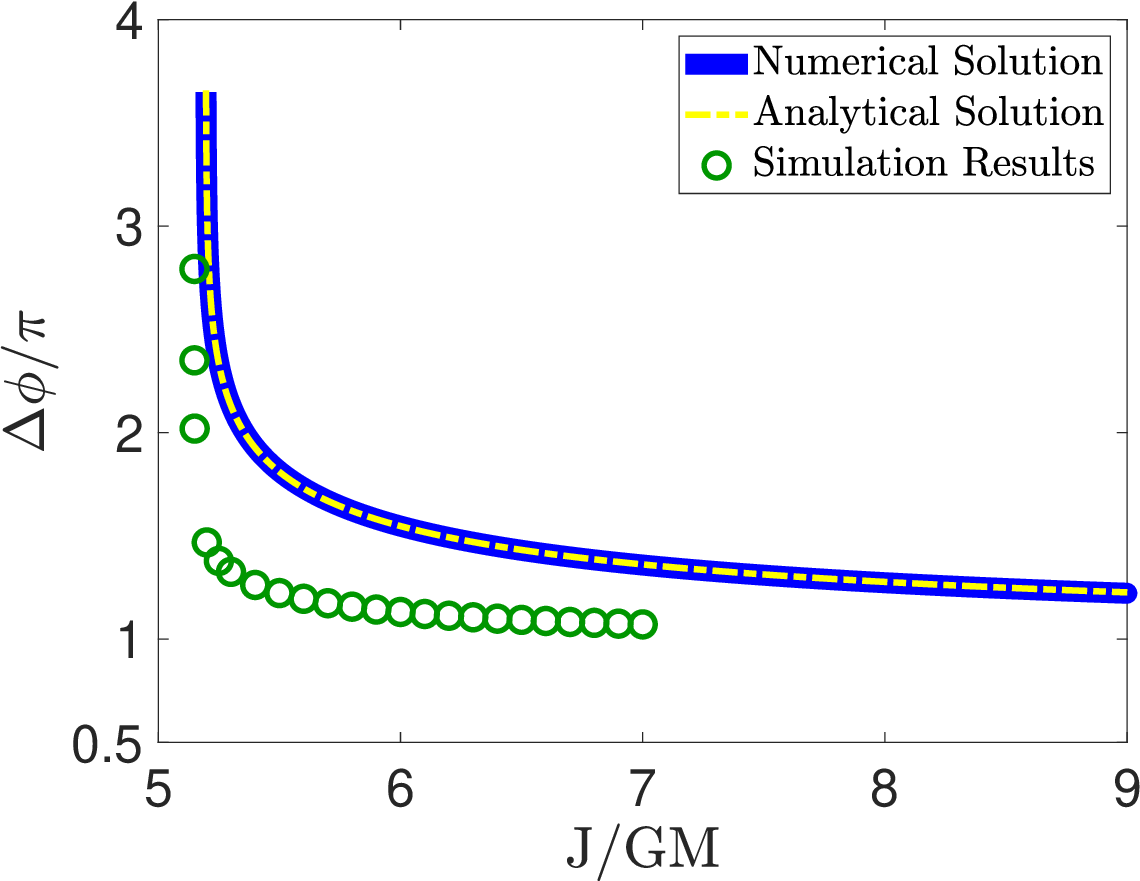}
\label{fig:deflect_angle_b}}
\\
\caption{Variation of pericentric distance $r_m$ (a) and deflection angle $\Delta\phi$ (b) as a function of $J/GM$. In both figures, 
blue line shows the result of analytical solution, and green circles come from simulated light orbits demonstrated in Fig.\ \ref{fig:deflect_orbits}. 
Yellow dash line in (b) represents numerical calculation of the integral for $\Delta\phi$.}
\label{fig:deflect_angle}
\end{figure}

Viewed from a distant observer, the geometry of spacetime plays the role of an effective potential that deviates light from its original path. 
To reveal the extent of deflection due to gravity, we first trace light orbits with different $J$ based on the E-L equation. 
As Fig.\ \ref{fig:deflect_orbits_a} shows, when light travels nearby the gravitation center, its trajectory is bent toward the center. 
The minimum of $r$, called the radial turning point and denoted by $r_m$, is reached at the critical condition that $dr/d\lambda=0$, which gives: 
\begin{equation}\label{eqn:rm_cubic}
\frac{r^3}{J^2}-r+2GM=0
\end{equation}

This is a cubic equation. From the Italian mathematician G.\ Cardano's formula in $15^{\text{th}}$ century, 
it is found that $r_m$ has a positive real solution only if the discriminant of 
eq.\ \eqref{eqn:rm_cubic} is negative: 
\begin{equation}
J^2 \sqrt{(GM)^2-\frac{J^2}{27}}\leq 0\ \Rightarrow\ J\geq 3\sqrt{3}GM
\end{equation}

The solution composes of two parts: 
\begin{equation}\label{eqn:Cardano}
\begin{split}
r_m&= r_1+r_2,\ \text{with} \\
r_{1,2}&= \left( -GMJ^2\pm i J^2 \sqrt{\frac{J^2}{27}-(GM)^2} \right)^{\frac{1}{3}}
\end{split}
\end{equation}

It is straightforward to check that $r_1 r_2=J^2/3$. The minimum of $r_m$ is taken at $J=3\sqrt{3}GM$, denoted by $J_m$, which equals: 
\begin{equation}
r_m=(-GM)^{\frac{1}{3}} J_m^{\frac{2}{3}}+\frac{\,1\,}{3} (-GM)^{-\frac{1}{3}} J_m^{\frac{4}{3}}=3GM
\end{equation}

which is just the critical radius of a circular orbit. Figs.\ \ref{fig:deflect_orbits_b}--\ref{fig:deflect_orbits_d} show different light orbits 
when $J$ takes values in the neighborhood of $J_m$. It is seen that, when $J$ approaches $J_m$, light tends to circulate around the medium center, 
forming a photon ring at $r=r_m$. 
In consequence, the deflection of light is greatly enhanced compared with normal gravitational deflection effect when $J\gg J_m$. 
Fig.\ \ref{fig:deflect_angle_a} shows the variation of $r_m$ as a function of $J/GM$. 
The blue line corresponds to the solution given by Cardano's formula in eq.\ \eqref{eqn:Cardano}, 
and the green circles represent simulated result, i.e.\ the radius of pericentric point on the light orbit. 
If $J$ is further decreased to be smaller than $J_m$, light will deviate inward from the circular orbit, falling inevitably into the center of the medium in spirals. 
In this case, $r_m$ loses its meaning and can no longer be defined. 

Due to the spherical symmetry of $n(r)$, light trajectories before and after reaching $r_m$ are mirrored to each other. 
Then the deflection angle $\Delta\phi$, which is the difference of $\phi$ between the incident and emergent directions, 
can be integrated as a function of $r$ or $u=1/r$ according to eq.\ \eqref{eqn:angle_element2}: 
\begin{equation}\label{eqn:deflect_integral}
\begin{split}
\Delta\phi&= 2\int_{r_m}^{+\infty} \frac{dr}{r \sqrt{\dfrac{r^2}{J^2}-\left( 1-\dfrac{2GM}{r} \right)}} \\
&= 2\int_{0}^{r_m^{-1}} \frac{du}{\sqrt{J^{-2}-u^2 \left( 1-2GMu \right)}}
\end{split}
\end{equation}

The major contribution to the integral comes from the interval near the end point $u=1/r_m$, in which the denominator is close to $0$. 
We thus devide the integration interval into two sections: $(0,\ r_m^{-1}-\epsilon)$ and $(r_m^{-1}-\epsilon,\ r_m^{-1})$, where $0<\epsilon\ll r_m^{-1}$, 
and introduce $v=1/r_m-u$ in the second section. The integral is calculated seperately in these two sections, and the result is: 
\begin{equation}\label{eqn:deflect_angle_a}
\begin{split}
\frac{\Delta\phi}{2}&= \int_{0}^{r_m^{-1}-\epsilon} \frac{du}{\sqrt{J^{-2}-u^2 \left( 1-GMu \right)^2}}
+\int_{0}^{\epsilon} \frac{dv}{\sqrt{\left( \dfrac{6GM}{r_m}-1 \right)v^2+\dfrac{2}{r_m} \left( 1-\dfrac{3GM}{r_m} \right)v}} \\
&= \arcsin J\xi+\frac{2GM}{J}-2GM \sqrt{\frac{1}{J^2}-\xi^2}+F(r_m,\epsilon)
\end{split}
\end{equation}

where we have defined: 
\begin{equation*}
\begin{split}
\xi&= \frac{r_m-GM}{r_m^2}+\left( \frac{2GM}{r_m}-1 \right) \epsilon, \\
F(r_m,\epsilon)&= \left\{
\renewcommand\arraystretch{2}
\begin{array}{l}
\dfrac{1}{\sqrt{6GM r_m^{-1}-1}} \text{arccosh} \left[ -\dfrac{r_m \left( r_m-6GM \right)}{r_m-3GM} \cdot \epsilon+1 \right]\ \left( 3GM<r_m<6GM \right) \\
\dfrac{1}{\sqrt{1-6GM r_m^{-1}}} \left\{ \dfrac{\pi}{2}-\arcsin \left[ -\dfrac{r_m \left( r_m-6GM \right)}{r_m-3GM} \cdot \epsilon+1 \right] \right\}\ \left( r_m>6GM \right)
\end{array}\right.
\end{split}
\end{equation*}

In the expression, $\epsilon$ is an arbitrarily small number, representing the extent to which $r\rightarrow r_m$. 
For the circular orbit itself, $r_m=3GM$ such that the factor in front of $\epsilon$ in the inverse hyperbolic cosine function goes to infinity. 
The integration diverges in this case, as light will continue circulating around the origin, with $\Delta\phi$ accumulating to infinity. 
For those orbits far away from $r_m$, the primary term in eq.\ \eqref{eqn:deflect_angle_a} is $2GM/J$, while other terms are negligibly small. 
Therefore the total deflection angle is $4GM/J$, a well known result in astronomical observations. 

In our calculation, we assign $0.3r_m^{-1}$ to $\epsilon$ in eq.\ \eqref{eqn:deflect_angle_a}. 
Besides the analytical formula, the integral of $\Delta\phi$ can be carried with high accuracy in numerical computation. 
In Fig.\ \ref{fig:deflect_angle_b}, we plot the results obtained in both ways (blue solid and yellow dash lines), together with the data 
corresponding to the simulated light orbits shown in Fig.\ \ref{fig:deflect_orbits} (green circles). 
As the graph shows, the blue and yellow dash lines nearly overlap, indicating the approximate degree of the analytical solution is excellent. 
The simulated result exhibits the same overall trend with moderate discrapencies from the analytical result. 
The error comes from the limit of simulation: $\Delta\phi$ and $J$ are read at a finite distance on the light orbit instead of infinitely far away. 
Consequently, they are both smaller than theoretical values, and thus the simulated data are all distributed at the bottom left of the blue line. 

The trend that smaller $J$ or $b$ results in larger $\Delta\phi$ is in agreement with the fact that light experiences stronger gravitational deflection effect 
if it passes by the gravitation center at a closer distance. It is worth noticing that, when $J$ decreases to the lower limit of $3\sqrt{3}GM$, 
the integral in eq.\ \eqref{eqn:deflect_angle_a} diverges and $\Delta\phi$ tends to approach infinity. 
In this case, light forms a circular orbit and such an orbit is analogous to the photon ring around a black hole \cite{photon_ring}. 

\begin{figure}[htp]
\subfigure[Light orbits in Newtonian limit.]{
\includegraphics[width=0.4\textwidth]{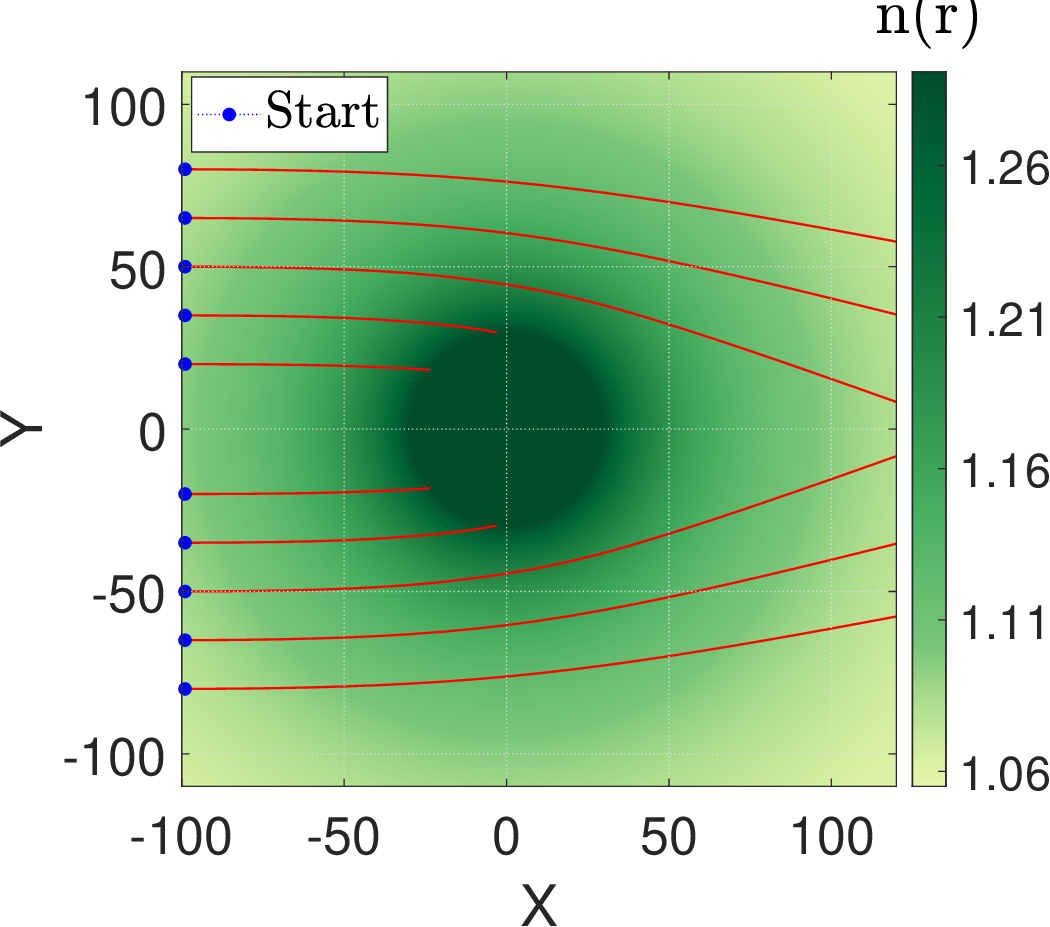}
\label{fig:Newton_limit_a}}
\quad
\subfigure[Deflection angle $\Delta\phi$ versus $J/GM$.]{
\includegraphics[width=0.43\textwidth]{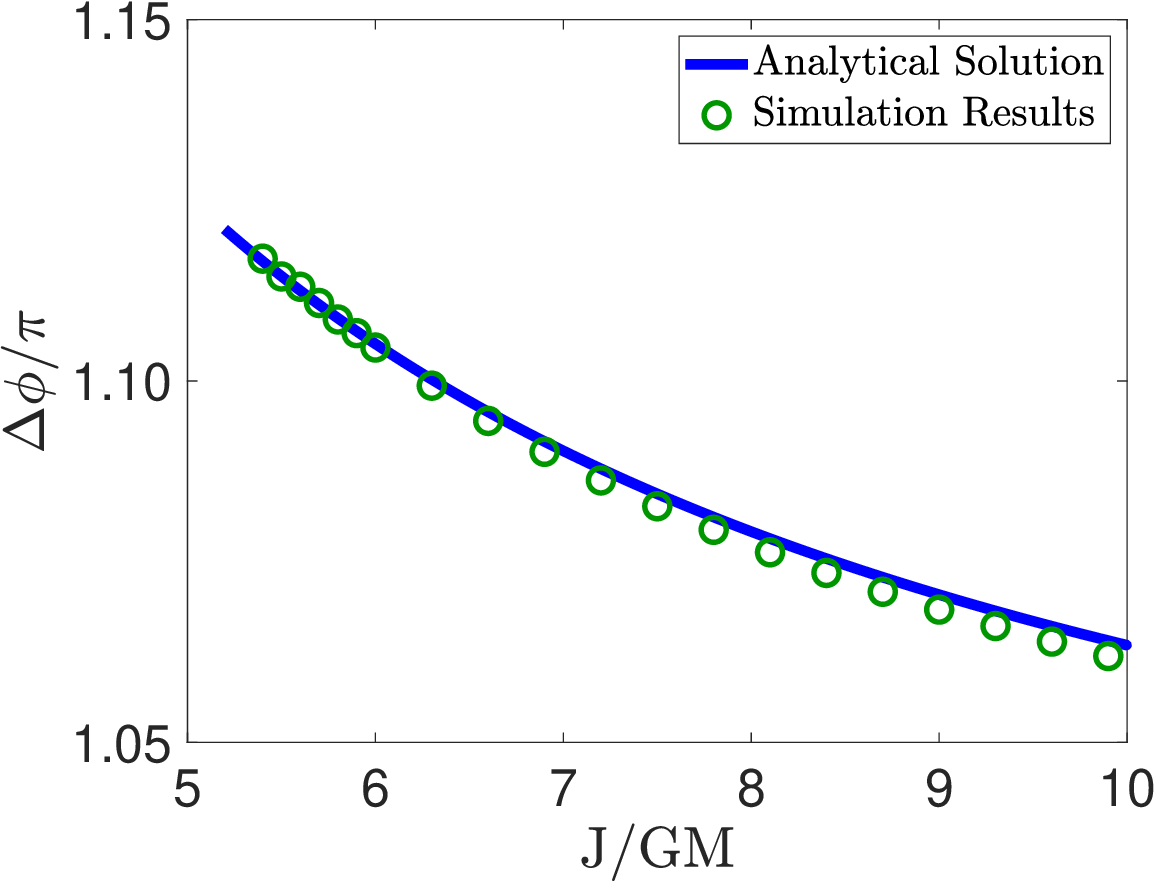}
\label{fig:Newton_limit_b}}
\\
\caption{(a) Light orbits with $n(\mathbi{r})$ taking the Newtonian limit for photons. The light rays are incident parallely from left to right. 
Depending on the distance from incident direction to the medium center, they are deflected to different extents. 
(b) Variation of $\Delta\phi$ versus $J/GM$, obtained both analytically (blue line) and numerically (green circles).}
\label{fig:Newton_limit}
\end{figure}

\section{Correspondance dictionary}
Through the above discussion, we have explored and established the connection between spacetime metric and refractive index profile of optical media. 
The analogical form is related to whether the rest mass of the test particle is zero or not. 
In tab.\ \ref{tab:dictionary} we summarize the corresponding quantities in details. 
In particular, when $g_{rr}\rightarrow 1$ and $GM/r\ll 1$, $n(r)$ approximates its Newtonian limit and so does the type of light orbit. 
The corresponding entries are also listed in the table. 

\begin{table}
\centering
\caption{Correspondance between kinematics in spacetime and refractive index in optics.}
\label{tab:dictionary}
\renewcommand\arraystretch{1.3}
\begin{tabular}{| p{5.2cm} p{4.7cm} p{6cm} |}
\hline
\textbf{quantities} & \textbf{particles} & \textbf{photons} \\
\hline
\hline
mass & $m$ & $0$ \\
geodesic parameter & $\tau$ & $\lambda$ \\
Lagrangian & $mv(\mathbi{r}) d\mathbi{r}$ & $cv^{-1}(\mathbi{r}) d\mathbi{r}$ \\
spatial velocity $v^2(\mathbi{r})$ & $-\dfrac{p_t^2}{g_{tt}}-1$ & 
$-\dfrac{g_{tt}}{g_{rr}} \left[ 1-\dfrac{p_\phi^2}{g_{rr} g_{\phi\phi}} \left( 1-\dfrac{g_{rr} r^2}{g_{\phi\phi}} \right) \right]$ \\
refractive index $n(\mathbi{r})$ & $\propto v(\mathbi{r})$ & $\propto v^{-1}(\mathbi{r})$ \\
$n(\mathbi{r})$ in Schwarzschild metric & $\left[ E^2 \left( 1-\dfrac{2GM}{r} \right)^{-1}-1 \right]^{\frac{1}{2}}$ & 
$\left( 1-\dfrac{2GM}{r} \right)^{-1} \left( 1+\dfrac{2GMJ^2}{r^3} \right)^{-\frac{1}{2}}$ \\
typical orbit & quasi-elliptical & asymptotically circular \\
$n(\mathbi{r})$ in Newtonian limit & $\left[ \dfrac{2GM}{r}-\dfrac{1-E^2}{E^2} \right]^{\frac{1}{2}}$ & 
$\left( 1+\dfrac{2GM}{r} \right)^{\frac{1}{2}}$ \\
typical orbit & elliptical & hyperbola-like \\
\hline
\end{tabular}
\end{table}

\subsection{Newtonian limit}
For massive particles, the profile of $n(r)$ in the Newtonian limit is simply the medium discussed in section \ref{sec:Newtonian spacetime}, 
in which light moves in elliptical orbits. For massless photons, we appoint $n(r)$ equal to the Newtonian limit in the E-L equation, 
and the resulting light orbits in numerical simulation are shown in Fig.\ \ref{fig:Newton_limit_a}. 
Obviously, the Newtonian approximation of $n(r)$ for particles and photons actually have similar forms, if we let $E>1$ in the expression of the former case. 
The astrophysical object with $E>1$ may be a comet coming from outside the solar system (for example, the famous intersteller object ‘Oumuamua \cite{Oumuamua}), 
so light moves along a hyperbola-like orbit similar to that of an alien comet, as we have seen in the simulated results. 
With simplified metric components, we rewrite eq.\ \eqref{eqn:angle_element2} in its Newtonian form as: 
\begin{equation}\label{eqn:angle_element3}
\frac{d\phi}{dr}=\frac{1}{r\sqrt{\dfrac{r^2}{J^2} \left( 1+\dfrac{2GM}{r} \right)-1}}
\end{equation}

At the pericentric radius $r_m$, the denominator equals zero, which gives the solution: 
\begin{equation}
r_m=\frac{J^2}{GM+\sqrt{(GM)^2+J^2}}=-GM+\sqrt{(GM)^2+J^2}
\end{equation}

Notice the discriminant in the solution is positive definite, with no restriction on the value of $J$. 
Then we carry the integration for $\Delta\phi$ on the interval $(r_m,+\infty)$, 
in a similar way as we deal with eq.\ \eqref{eqn:deflect_integral}, and the result is: 
\begin{equation}\label{eqn:deflect_angle_b}
\begin{split}
\Delta\phi&= 2\int_{r_m}^{+\infty} \frac{dr}{r \sqrt{\dfrac{r^2}{J^2} \left( 1+\dfrac{2GM}{r} \right)-1}} \\
&= 2\int_{0}^{r_m^{-1}} \frac{b du}{\sqrt{-J^2 u^2+2GMu+1}} \\
&= \pi+2\arcsin \left[ \left( \frac{J}{GM} \right)^2+1 \right]^{-\frac{1}{2}}
\end{split}
\end{equation}

Based on eq.\ \eqref{eqn:deflect_angle_b}, we plot the dependence of $\Delta\phi$ on $J/GM$ in Fig.\ \ref{fig:Newton_limit_b}, 
together with the results calculated numerically for different light orbits. It is found that the analytical and simulated results agree well with each other. 
Moreover, $\Delta\phi$ gets closer to the value in Schwarzschild spacetime at larger $J$, as revealed in comparison with Fig.\ \ref{fig:deflect_angle_b}. 
This is a natural result considering that the spacetime geometry transits gradually to the Newtonian limit when $r$ increases large enough. 
However, error is always present. Since eq.\ \eqref{eqn:deflect_angle_b} can be considered to equal $\pi+2GM/J$ when $J/GM\gg 1$, 
light rays will deviate from their original paths by $2GM/J$, exactly half of that given by eq.\ \eqref{eqn:deflect_angle_a} for distant rays. 
Toward smaller $J$, $\Delta\phi$ predicted by eq.\ \eqref{eqn:deflect_angle_b} tends to reach a maximum of $2\pi$. 
It deviates significantly from the full result in eq.\ \eqref{eqn:deflect_angle_a}, which diverges instead, since the Newtonian approximation fails in this situation. 

\subsection{Further simplification}
If we further simplify $n(r)$ to approximately equal $1+GM/r$, 
the E-L equation will take a relatively simple form in this case, which can be solved analytically. The solution is: 
\begin{equation}
\begin{split}
\phi&= \frac{K}{\sqrt{K^2-(GM)^2}}
\arccos\left[ \frac{K-GM}{\left( \dfrac{1}{K}+\dfrac{GM}{K^2} \right) \left( r+GM \right)^2-2GM} \right]^{\frac{1}{2}}+\phi_0 \\
\text{where}&\ K:=J+GM\sin\phi_i
\end{split}
\end{equation}

Here $K$ is defined for the light ray with angular momentum $J$, entering the medium horizontally at the point $(r_i,\phi_i)$. 
For a specific light ray, $K$ is a constant that happens to equal the Fermat's ray invariant defined for a spherically symmetric GRIN lens \cite{Fermat_invariant}. 
Then we obtain the expression for the deflection angle: 
\begin{equation}\label{eqn:deflect_angle_c}
\Delta\phi=\frac{\pi}{\sqrt{1-\left( \dfrac{GM}{K} \right)^2}}=\frac{\pi}{\sqrt{1-\left( \dfrac{J}{GM}+\sin\phi_i \right)^{-2}}}
\end{equation}

It is found that compared to the analytical solution in eq.\ \eqref{eqn:deflect_angle_b}, as shown in Fig.\ \ref{fig:Newton_limit_b}, 
the output of eq.\ \eqref{eqn:deflect_angle_c} is smaller over the whole range of $J$, 
indicating that the ss $n(r)$ can only be used to predict the trend but not the exact value of the deflection angle. 
At the same time, like the solutions in other conditions, eq.\ \eqref{eqn:deflect_angle_c} converges to $\pi$ as $J$ increases, 
reducing to the case in which light rays are far away from the gravitational attraction center and hardly deflected by the curved spacetime. 
The trajectories are just straight lines, so $\Delta\phi$ is nearly equal to $\pi$. 

Using the spacetime metric and refractive index dictionary, we can design the corresponding optical medium in which light orbit coincides with the trajectory 
of a general particle in gravitational fields. This method applies not only to particles with nonzero rest mass, but also to photons with no mass. 
By connecting spacetime geometry to basic optical constant like the refractive index, 
we can use the refractive index distribution of optical media to simulate gravitational fields, 
hence making it possible to study astrophysical problems in artificial optical media.

\section{Conclusion}
In summary, we relate the refractive index of optical media to the geometry of spacetime in gravitational fields from the analogy between mechanics and optics, 
in view of the least action principle they have in common. 
This work extends $F=ma$ optics to curved spacetime in general relativity, where the gravity has been geometrized. 
It suggests a way of realizing the intrinsic geometry of spacetime in an optical medium, as long as its refractive index profile reflects the velocity distribution of 
free falling particles, which is then linked to the metric components. 
The wide variety of motion types of cosmic bodies brings abundance into optical design through this connection. 
In reverse, these results provide a reference for the study of optical problems in cosmology. 
Given the propagation path of light in outer space, we can duplicate that path via designing $n(\mathbi{r})$ and the corresponding optical materials. 
In this manner, we can gain more information about the vast universe in our laboratories, 
as if we are bending and distorting space according to practical needs.

\bibliographystyle{ieeetr}
\bibliography{bibliolist}

\end{document}